\documentclass[12pt]{article}
\usepackage{amsfonts}
\usepackage{amsmath}
\usepackage{amssymb}
\usepackage{amsthm}
\usepackage{graphicx}
\usepackage{latexsym}
\usepackage{xy}
\usepackage{verbatim}
\usepackage{bm}
\theoremstyle{plain}

\theoremstyle{definition}
\newtheorem{definition}{Definition}[section]

\theoremstyle{remark}

\newtheorem*{pn}{Proof}

\newcommand{\non}{\nonumber}

\newcommand{\bd}{\begin{definition}}
\newcommand{\ed}{\end{definition}}
\newcommand{\bp}{\begin{pn}}
\newcommand{\ep}{\end{pn}}

\numberwithin{equation}{section}

\parskip=\baselineskip

\begin{document}

\begin{titlepage}
\begin{flushright}

\end{flushright}
\vskip 1.5in
\begin{center}
{\bf\Large{Non-Supersymmetric Attractor Flow in Symmetric Spaces}}
\vskip 0.5in {Davide Gaiotto, Wei Li and Megha Padi} \vskip 0.3in
{\small{ \textit{ Jefferson Physical Laboratory, Harvard
University, Cambridge MA 02138, USA}}}

\end{center}
\vskip 0.5in

\baselineskip 16pt
\date{}

\begin{abstract}
We derive extremal black hole solutions for a variety of four
dimensional models which, after Kaluza-Klein reduction, admit a
description in terms of 3D gravity coupled to a sigma model with
symmetric target space. The solutions are in correspondence with
certain nilpotent generators of the isometry group. In particular,
we provide the exact solution for a non-BPS black hole with
generic charges and asymptotic moduli in ${\cal N}=2$ supergravity
coupled to one vector multiplet. Multi-centered solutions can also
be generated with this technique. It is shown that the non-supersymmetric
solutions lack the intricate moduli space of bound configurations
that are typical of the supersymmetric case.

\end{abstract}
\end{titlepage}
\vfill\eject

\tableofcontents
\section{Introduction}

Soon after the attractor mechanism was first discovered in
supersymmetric (BPS) black holes \cite{Ferrara:1995ih}, it was
reformulated in terms of motion on an effective potential for the
moduli \cite{Ferrara:1997tw}. Ferrara et al demonstrated that the
critical points of this potential correspond to the attractor
values of the moduli. More recently, several groups used the
effective potential to show that non-supersymmetric (non-BPS)
extremal black holes can also exhibit the attractor mechanism,
thereby creating a new and exciting field of research
\cite{Kallosh:2006bt, Goldstein:2005hq}. Many connections between
non-BPS attractors and other active areas of string theory soon
revealed themselves. Andrianopoli et al found that both BPS and non-BPS
black holes embedded in a supergravity with a symmetric moduli
space can be studied using the same formalism, and they uncovered
many intricate relations between the two \cite{Andrianopoli:2006ub, D'Auria:2007ev}.
Dabholkar, Sen and Trivedi proposed a microstate counting for
non-BPS black holes (albeit subject to certain constraints
\cite{Dabholkar:2006tb}). Saraikin and Vafa suggested that a new
extension of topological string theory generalizes the
Ooguri-Strominger-Vafa (OSV) formula such that it is also valid
for non-supersymmetric black holes \cite{Saraikin:2007jc}.
Studying non-BPS attractors could also give insight into
non-supersymmetric flux vacua. Given all these possible
applications, it is important to characterize non-BPS black holes
as fully as possible.

There has been a great deal of progress in understanding the
near-horizon region of these non-BPS attractors. The second
derivative of the effective potential at the critical point
determines whether the black hole is an attractor, and the
location of the critical point yields the values of the moduli at
the horizon; in this way, one can compute the stability and
attractor moduli for all models with cubic prepotential
\cite{Tripathy:2005qp,Nampuri:2007gv}. However, the effective
potential has only been formulated for the leading-order terms in
the supergravity lagrangian. If one wants to include
higher-derivative corrections, one can instead use Sen's entropy
formalism, which incorporates Wald's formula, to characterize the
near-horizon geometry in greater generality \cite{Sen:2005wa}.
Sen's method has led to many new results \cite{Sahoo:2006pm,
Astefanesei:2006dd, Sahoo:2006rp, Chandrasekhar:2006kx,
Alishahiha:2006ke, Sen:2005iz}.  The tradeoff is that this method
cannot be used to determine any properties of the solution away
from the horizon.

The BPS attractor flow is constructed from the attractor value
$z^{*}_{BPS}=z^{*}_{BPS}(p^I,q_I)$ by simply replacing the D-brane
charges with the corresponding harmonic functions:
\begin{equation}
z_{BPS}(\vec{x})=z^{*}_{BPS}(p^I \rightarrow H^I(\vec{x}),q_I
\rightarrow H_I(\vec{x}))
\end{equation}
where the harmonic functions are
\begin{displaymath}
\left(%
\begin{array}{c}
  H^I(\vec{x}) \\
  H_I(\vec{x}) \\
\end{array}%
\right)=\left(%
\begin{array}{c}
  h^I \\
  h_I \\
\end{array}%
\right)+\frac{1}{|\vec{x}|}\left(%
\begin{array}{c}
  p^I \\
  q_I \\
\end{array}%
\right)
\end{displaymath}
Moreover, this procedure can be applied to construct the
multi-centered BPS attractor flow that describes the
supersymmetric black hole bound state \cite{Bates:2003vx}, where
the harmonic functions are generalized to have multiple centers:
\begin{displaymath}
\left(%
\begin{array}{c}
  H^I(\vec{x}) \\
  H_I(\vec{x}) \\
\end{array}%
\right)=\left(%
\begin{array}{c}
  h^I \\
  h_I \\
\end{array}%
\right)+\sum_{i}\frac{1}{|\vec{x}-\vec{x}_i|}\left(%
\begin{array}{c}
  (p^I)_i \\
  (q_I)_i \\
\end{array}%
\right)
\end{displaymath}
It is conjectured in \cite{Kallosh:2006ib} that the non-BPS flow
can be generated in the same fashion, namely, by replacing the
charges in the attractor value with the corresponding harmonic
functions. However, as will be proven in this paper, this
procedure does not work for systems with generic charge and
asymptotic moduli.\footnote{This has also been shown in \cite{Lopes Cardoso:2007ky}.
}

In principle, one could construct the full non-BPS flow (the black
hole metric, together with the attractor flow of the moduli) by
solving the equation of motion derived from the lagrangian.
However, this is a second-order differential equation and only
reduces to a first-order equation upon demanding the preservation
of supersymmetry. Ceresole
et al have written down an equivalent first-order equation in terms of a ``fake superpotential'', but so far, the fake superpotential can only be explicitly constructed for special charges and asymptotic moduli \cite{Ceresole:2007wx, Lopes Cardoso:2007ky}. The
most generic non-BPS equation of motion is complicated enough that
it has not yet been solved. Similarly, multi-centered non-BPS
black holes have not been studied.

Our goal is to construct the full flow for non-BPS
stationary black holes in four dimensions. Instead of directly solving the equation of motion,
we reduce the action on the timelike isometry and dualize all 4D
vectors to scalars. The new moduli space ${\cal M}_{3D}$ contains
isometries corresponding to all the charges of the black hole, and
the black hole solutions are simply geodesics on ${\cal M}_{3D}$.
This method was introduced in \cite{Breitenlohner:1987dg} and has been used to construct static and rotating black holes in
heterotic string theory \cite{Cvetic:1995kv, Cvetic:1996xz} and to study the classical BPS single-centered flow and its radial
quantization \cite{Gunaydin:2005mx,
Neitzke:2007ke}.

In this paper, we work in two specific theories of gravity, but we
expect that this method can be used for any model whose ${\cal
M}_{3D}$ is symmetric. The basic technique is reviewed in more
detail in Section 2.  In Section 3, we show how this method works
in a simple case: the toroidal compactification of $D$-dimensional
pure gravity. Section 4 serves as an introduction to
single-centered attractor flow in ${\cal N}=2$ supergravity
coupled to one vector multiplet, and Sections 5 and 6 are
dedicated to constructing the full flows for both BPS and non-BPS
single-centered black holes with generic charges. We find that
they are generated by the action of different classes of nilpotent
elements in the coset algebra. Both types of flows are shown to
reach the correct attractor values at the horizon. In Section 7,
the procedure is generalized straightforwardly to construct both
BPS and non-BPS multi-centered solutions. We use a metric ansatz
with a flat spatial slice and we are able to recover the BPS bound
states described by Bates and Denef. Using the same ansatz, we are
able to build non-BPS multi-centered solutions. Unfortunately,
solutions generated this way turn out to always have charges at
each center which are mutually local. The last section reviews our
conclusions and suggests possibilities for future work.

\section{Framework}

Here we outline the method we will use to construct black hole solutions. This method was first described in \cite{Breitenlohner:1987dg}. We first reduce a general gravity action from four dimensions down to three, and derive the equation of motion. We then specialize to certain theories which have a 3d description in terms of a symmetric coset space. In such situations, we can easily find solutions to the equation of motion. The solutions are geodesics (or generalizations thereof) on the 3d moduli space and they are generated by elements of the coset algebra.

\subsection{3D Moduli Space}

We will study stationary solutions in a theory with gravity coupled to scalar and vector matter. Let the scalars be $z^i$ and the vectors be $A^I$. Then the most general ansatz for a
stationary solution in four dimensions is:
\begin{eqnarray}
ds^2 &=&-e^{2U} (dt+\omega)^2+e^{-2U} \textbf{g}_{ab}dx^a dx^b \\
F^I&=&dA^I=d\left(A^I_0(dt+\omega)+\mathbf{A}^I \right)
\end{eqnarray}
where $a,b=1,2,3$ label the spatial directions and bold font
denotes three-dimensional fields and operators.  Since none of the
fields are time-dependent, we can compactify on the time isometry
and reduce to three-dimensional space $\mathcal{M}_{3D}$. This
procedure is called the $c^{*}$-map.  In three dimensions, a
vector is Hodge-dual to a scalar. The equations of motion for
$\omega$ and the gauge fields allow us to define the dual scalars
$\phi_{\omega}$ and $\phi_{\mathbf{A}^I}$.

We then obtain the 3d lagrangian in terms of only scalars
\begin{equation}
\mathcal{L} = \frac{1}{2}\sqrt{\textbf{g}} (
-\frac{1}{\kappa}\textbf{R}+\partial_a \phi^m
\partial^a \phi^ng_{mn})
\end{equation}
where $\phi^n$ are the moduli fields
\begin{equation}
\phi^n=\{U,z^i,\bar{z}^{\bar{i}},\phi_{\omega},A_0^I,\phi_{\mathbf{A}^I}\}
\end{equation}
and $\textbf{g}_{ab}$ is the space time metric and $g_{mn}$ is the
metric of a manifold $\mathcal{M}_{3D}$. The system is $3D$
gravity minimally coupled to a nonlinear sigma model with moduli
space $\mathcal{M}_{3D}$. Next, we will find the equation of
motion in this theory.

\subsection{Attractor Flow Equation}

The equation of motion of $3D$ gravity is Einstein's equation:
\begin{equation}
\textbf{R}_{ab}-\frac{1}{2}\textbf{g}_{ab} \textbf{R}=\kappa
T_{ab}=\kappa (
\partial_a \phi^m
\partial_b \phi^ng_{mn}-\frac{1}{2}\textbf{g}_{ab}\partial_c \phi^m
\partial^c \phi^ng_{mn})
\end{equation}
and the equation of motion of the moduli is:
\begin{equation}
\nabla_a \nabla^a \phi^n + \Gamma^{n}_{mp} \partial_a \phi^m
\partial^a
\phi^p=0
\end{equation}
For simplicity, we consider only the case where the $3D$ spatial
slice is flat (it is guaranteed to be flat only for extremal
single-centered black holes). Then the dynamics of the moduli are
decoupled from that of the $3D$ gravity:
\begin{equation}
\textbf{R}_{ab}= 0 \qquad \Longrightarrow \qquad
\partial_a \phi^m  \partial_b \phi^n g_{mn}=0
\end{equation}
  In the multi-centered case, we need to solve the full equations for the
moduli as functions of the 3d coordinates $\vec{x}$. For
single-centered solutions, the moduli only depend on $r$; to
satisfy the above equations, the motion of the moduli must follow
null geodesics inside $\mathcal{M}_{3D}$.

A generic null geodesic flows to the boundary of the moduli space
$\mathcal{M}_{3D}$. A single-centered attractor flow is defined as
a null geodesic that terminates at a point on the $U \rightarrow
-\infty$ boundary and in the interior region with respect to all
other coordinates. This is guaranteed for the BPS attractor by the
constraints imposed by the supersymmetry. To find the
single-centered non-supersymmetric attractor flow, one needs to
find a way to construct null geodesics and a constraint that pick
out the ones that stop at this specific component of the boundary.
In the next section, we will show that we can do this for models
with special properties, and the method can be easily generalized
to find the multi-centered attractor solution.

\subsection{Models with Symmetric Moduli Space}

The problem of finding such a constraint in a generic model is not
easy. To simplify, we study any model whose moduli space is a
symmetric homogeneous space:
$\mathcal{M}_{3D}=\mathbf{G}/\mathbf{H}$. When $\mathcal{M}_{3D}$
is a homogeneous space, the isometry group $\mathbf{G}$ acts
transitively on $\mathcal{M}_{3D}$. $\mathbf{H}$ denotes the
isotropy group, which is the maximal compact subgroup of
$\mathbf{G}$ when one compactifies on a spatial isometry down to
$(1,2)$ space, or the analytical continuation of the maximal
compact subgroup of $\mathbf{G}$ when one compactifies on the time
isometry down to $(0,3)$ space. The Lie algebra $\mathbf{g}$ has
the Cartan decomposition: $\mathbf{g}=\mathbf{h}\oplus \mathbf{k}$
where
\begin{equation}
[\mathbf{h},\mathbf{h}]=\mathbf{h} \qquad
[\mathbf{h},\mathbf{k}]=\mathbf{k}
\end{equation}
When $\mathbf{G}$ is semi-simple, the homogeneous space is
symmetric, and
\begin{equation}
[\mathbf{k},\mathbf{k}]=\mathbf{h}
\end{equation}

The models with symmetric moduli space includes: $D$-dimensional
gravity toroidally compactified to four dimensions, certain models
of 4D $\mathcal{N}=2$ supergravity coupled to vector-multiplet,
and all 4D ${\cal N}
> 2$ extended supergravity. The entropy of the last two classes is U-duality invariant. In the present paper, we will only
consider the first two classes, namely, the $D$-dimensional
gravity toroidally compactified to four dimensions, and the 4D
$\mathcal{N}=2$ supergravity coupled to $n_V$ vector-multiplet.

The left-invariant current is
\begin{equation}
J=M^{-1}dM=J_{\mathbf{k}}+J_{\mathbf{h}}
\end{equation}
where $M$ is the coset representative, and $J_{\mathbf{k}}$ is the
projection of $J$ onto the coset algebra $\mathbf{k}$. The
lagrangian density of the sigma-model with target space
$\mathbf{G}/\mathbf{H}$ is given by $J_{\mathbf{k}}$ as:
\begin{equation}
L=\hbox{Tr}(J_{\mathbf{k}}\wedge *_{3} J_{\mathbf{k}})
\end{equation}

The geodesic of the homogeneous space written in terms of the
coset representative is simply
\begin{equation}
M=M_0e^{k \tau/2}\qquad \textrm{with} \qquad k \in \mathbf{k}
\end{equation}
where $M_0$ parameterizes the initial point, and the $\frac{1}{2}$
is for later convenience. A null geodesic has zero length:
\begin{equation}
|k|^2=0
\end{equation}
Therefore, in a homogeneous space, we can find the null geodesics
that end at an attractor point by imposing the appropriate
constraint on the null elements of the coset algebra.

Since $M$ is defined up to the action of the isotropy group
$\mathbf{H}$, in order to read off the moduli fields from $M$ in
an $\mathbf{H}$-independent way, we construct the symmetric matrix
using the metric signature matrix $S_0$:
\begin{equation}
S \equiv MS_0 M^T
\end{equation}
In all systems considered in the present paper, $\mathbf{H}$ is
the maximal orthogonal subgroup of $\mathbf{G}$ with the correct
signature:
\begin{equation}
HS_0H^T=S_0 \qquad \textrm{for}\,\,\, \forall H \in \mathbf{H}
\end{equation}
That is, the isotropy group $\mathbf{H}$ preserves the symmetric
metric matrix $S_0$. Therefore, $S$ is invariant under
$M\rightarrow MH$ with $H \in \mathbf{H}$. Moreover, as the
isotropy group $\mathbf{H}$ acts transitively on the space of of
matrices with a given signature, the space of possible $S$ is the
same as the symmetric space $\mathbf{G}/\mathbf{H}$. That is, the
moduli of $\mathbf{G}/\mathbf{H}$ can be combined into the
symmetric matrix $S$. And the current of $S$ is
\begin{equation}
J_S=S^{-1}dS
\end{equation}

It is easy to perform the projection onto the coset algebra
$\mathbf{k}$. The (generalized) orthogonality condition of the
isotropy group $\mathbf{H}$ can be expressed in terms of the
subalgebra element $h$ which is in $H=e^{h}$ as
\begin{equation}
hS_0+S_0h^T=0 \qquad \qquad \forall \, h\in \mathbf{h}
\end{equation}
In other words, $(\mathbf{h}S_0)$ is anti-symmetric:
$(\mathbf{h}S_0)^T=-(\mathbf{h}S_0)$. Thus the coset algebra,
being the compliment of $\mathbf{h}$, can be defined as the
$\mathbf{k}$ with $(\mathbf{k}S_0)$ being symmetric:
$(\mathbf{k}S_0)^T=(\mathbf{k}S_0)$, i.e.
\begin{equation}
k^T=S^{-1}_0kS_0 \qquad \qquad \forall \, k\in \mathbf{k}
\end{equation}
Therefore, the projection of an element $g$ in $\mathbf{g}$ onto
the coset algebra $\mathbf{k}$ is:
\begin{equation}
g_{\mathbf{k}}=\frac{g+S_0 g^T S^{-1}_0}{2}
\end{equation}
For the left-invariant current $J=M^{-1}dM$, the projection onto
$\mathbf{k}$ is:
\begin{equation}
J_{\mathbf{k}}=\frac{J+S_0 J^T S^{-1}_0}{2}
\end{equation}

It is straightforward to show that the current constructed from
$S$ is related to the projected left-invariant current
$J_{\mathbf{k}}$ by:
\begin{equation}
J_{S}=S^{-1}dS=2(S_0M^T)^{-1}J_{\mathbf{k}}(S_0M^T)
\end{equation}
The lagrangian in terms of $S$ is thus $
L=\frac{1}{4}\hbox{Tr}(J_{S}\wedge *_{3} J_{S})$. That is, the
lagrangian density is
\begin{equation}
{\mathcal L} = \frac{1}{4}\hbox{Tr} (S^{-1}\nabla S \cdot
S^{-1}\nabla S )
\end{equation}
which is invariant under the action of the isometry group
$\mathbf{G}$:
\begin{equation}
S \rightarrow G^{-1}S G \qquad \textrm{where} \qquad G \in \mathbf{G}
\end{equation}
and whose conserved current is:
\begin{equation}
J=S^{-1} \nabla S
\end{equation}
where we have dropped the subscript $S$ in $J_S$, since we will
only be dealing with this current from now on. The equation of
motion is the conservation of the current:
\begin{equation}
\nabla \cdot J=\nabla \cdot (S^{-1} \nabla S)=0
\end{equation}

We now specialize to the single-centered solutions: they correspond
to geodesics in the coset manifold. The spherical symmetry allows
the $3d$ metric to be parameterized as
\begin{equation}
ds^2_3=C(r)^2 d \vec x^2
\end{equation}
 Then the equations of
motion involve the operator $d_r r^2 C(r) d_r$, and reduce to
geodesic equations in terms of a parameter $\tau$ such that
\begin{equation}
\frac{dr}{d\tau} = r^2 C(r) \ .
\end{equation} The function $C(r)$ is then
determined from the equations of motion of $3d$ gravity. The
equations of motion can be written as
\begin{equation}
\frac{d}{d\tau}(S^{-1}\frac{dS}{d\tau} )= 0
\end{equation}
In the extremal limit the geodesics become null, the $3d$ metric
is flat and
\begin{equation}
\tau= - \frac{1}{r}
\end{equation}

In the search for multi-centered extremal solutions, where the
spherical symmetry is absent, it is very convenient to restrict to
solutions with a flat 3d metric. This is consistent with the
equations of motion as long as the $3d$ energy momentum tensor  is
zero everywhere:
\begin{equation}
T_{ab}=Tr (J_a J_b)=0
\end{equation} The coupled problem with generic non-flat $3d$ metric
is much harder, and exact solutions are hard to find unless a
second Killing vector is present.

Since different values of the scalars at infinity are easily
obtained by a $\mathbf{G}$ transformation, to start with, we will
consider the flow starting from $M_0=1$, and generalize to generic
asymptotic moduli later. For a single-centered solution, the flow
of $M$ is $M=M_0e^{k\tau/2}$. Since all the coset representatives
can be brought into the form $e^{g}$ with some $g\in\mathbf{k}$ by
an $\mathbf{H}$-action, we can write $M_0=e^{g/2}$, so
$M=e^{g/2}e^{k\tau/2}$. And the flow of $S$ is
\begin{equation}\label{Stauarbitrymoduli}
S(\tau)=e^{g/2}e^{k\tau}e^{g/2}S_0
\end{equation} The charges of the solution are read from the conserved
currents
\begin{equation}
J(r)=S^{-1} \nabla S
=\frac{S_0e^{-g/2}ke^{g/2}S_0}{r^2}\hat{\vec{r}}
\end{equation}

\section{Torus Reduction of $D$-dimensional Pure Gravity}

Now we use the method introduced in the previous section to analyze pure gravity toroidally compactified down to four dimensions. We explain why the attractor flow generator, $k$, needs to be nilpotent, and we find the Jordan forms of $k^2$ and $k$. Using this information, we construct single-centered attractor flows. We then generalize to multicentered black holes in pure gravity and show that these solutions have mutually local charges and no intrinsic angular momentum.

\subsection{Kaluza-Klein Reduction}

The simplest example of a system that admits a $3d$ description in
terms of a sigma model on a symmetric space is pure gravity in $D$
dimensions, compactified on a $D-4$ torus. The KK reduction to
$4D$ parameterizes the metric as
\begin{equation}
ds^2_D=\rho_{pq}(dy^p + A_\mu^p dx^\mu)(dy^q + A_\nu^q dx^\nu)+
\frac{1}{ \sqrt{\det \rho}} ds^2_4 \quad \quad 1 \leq p,q \leq D-4
\end{equation}
Here $y^p$ are the torus coordinates, $x^\mu$ the coordinates on
$R^{3,1}$, and $\rho_{pq}$ the metric of the torus. A $4D$ metric
with one timelike Killing spinor is then parameterized as
\begin{equation}
ds^2_4=- u (dt + \omega_i dx^i)^2+ \frac{1}{u} ds^2_3
\end{equation}
where $u=e^{2U}$, to connect with the parametrization in the later
part of the paper; and $i=1,2,3$ denote the $3d$ space coordinates.

The two expressions combine as
\begin{equation}
ds^2_D=G_{ab}(dy^a + \tilde{\omega}_i^a dx^i)(dy^b +
\tilde{\omega}_j^b dx^j)+ \frac{1}{- \det G} ds^2_3 \quad \quad 0
\leq a,b \leq D-4
\end{equation}
Here $y^a$ are the torus coordinates plus time, $x^i$ coordinates
on $R^3$ and
\begin{equation}
G=\left(\begin{array}{cc} \rho_{pq} & \rho_{pr}A_0^r \\
A_0^r\rho_{rq} & A_0^r\rho_{rs} A_0^s - \frac{u}{ \sqrt{\det
\rho}}
\end{array} \right)
\end{equation}
and $\tilde{\omega}^{a}=(\tilde{\omega}^{p},\tilde{\omega}^{0})$
is:
\begin{equation}
\tilde{\omega}^p = (A_i^p - A_0^p \omega_i)dx^i \qquad \qquad
\tilde{\omega}^0 = \omega
\end{equation}

If the forms $\tilde{\omega}^a$ are dualized to scalars $\alpha_a$
as
\begin{equation}
d\alpha_a = -\det G G_{ab} *_3 d\tilde{\omega}^b
\end{equation}
the various scalars can be combined into a symmetric unimodular
$(D-2) \times (D-2)$ matrix
\begin{equation}
S=\left(\begin{array}{cc} G_{ab} + \frac{1}{\det G} \alpha_a \alpha_b & \frac{1}{\det G} \alpha_a \\ \frac{1}{\det G} \alpha_b & \frac{1}{\det G} \end{array} \right)
\end{equation}
In terms of the $4D$ fields that is
\begin{equation}
S=\left(\begin{array}{ccc} \rho_{pq} - \frac{1}{u \sqrt{\det
\rho}} \alpha_p \alpha_q & \rho_{pr}A_0^r - \frac{1}{u \sqrt{\det
\rho}} \alpha_p \alpha_0
& -\frac{1}{u \sqrt{\det \rho}} \alpha_p \\
A_0^r\rho_{rq} - \frac{1}{u \sqrt{\det \rho}} \alpha_0 \alpha_q
& A_0^r\rho_{rs} A_0^s - \frac{u}{ \sqrt{\det \rho}}- \frac{1}{u \sqrt{\det \rho}} \alpha_0 \alpha_0 & -\frac{1}{u \sqrt{\det \rho}} \alpha_0 \\
-\frac{1}{u \sqrt{\det \rho}} \alpha_q & -\frac{1}{u \sqrt{\det
\rho}} \alpha_0 & -\frac{1}{u \sqrt{\det \rho}} \end{array}
\right)
\end{equation}

The equations of motion derive from the lagrangian density
$\mathcal{L}=Tr \nabla S S^{-1} \nabla S S^{-1}$, invariant under
$S \to U^T S U$ for any $U$ in $SL(D-2)$. As this $SL(D-2)$ action
is transitive on the space of matrices with a given signature, the
space of possible $S$ is the same as the symmetric space
$SL(D-2)/SO(D-4,2)$. Notice that the signature of the stabilizer
$SO(D-4,2)$ is appropriate for the reduction from $(D-1,1)$ to
$(3,0)$ signature. The usual reduction from $(D-1,1)$ to $(2,1)$
would give a $SL(D-2)/SO(D-2)$, while the Euclidean reduction from
$(D,0)$ to $(3,0)$ gives $SL(D-2)/SO(D-3,1)$ \cite{Hull:1998br}.

The coset representative under the left $SO(D-4,2)$ action can be
described in terms of a set of vielbeins
\begin{equation}
e^A = E^A_a (dy^a + \omega_i^a dx^i) \qquad \qquad e^I =
\frac{1}{\det M} e^I_{(3)}
\end{equation}
as
\begin{equation}
M=\left(\begin{array}{cc} E^A_a & 0 \\ \frac{1}{\det E} \alpha_b &
\frac{1}{\det E} \end{array} \right)
\end{equation}
Then the symmetric $SO(D-4,2)$ invariant matrix
\begin{equation}S
= MS_0 M^T
\end{equation}
can be used to read off the solution more easily. Without loss of
generality, we can take $S_0$ to be the signature matrix:
\begin{equation}
S_0=Diag(\eta,-1)=Diag(1,\cdots,1,-1,-1)
\end{equation}

The equations of motion are equivalent to the conservation of the
$SL(D-2)$ currents $J=S^{-1} dS$. Some of those currents
correspond to the usual gauge currents in 4D: the first $D-4$
elements of the last column $J_{i,D-2}$ are the KK monopole
currents, the first $D-4$ elements of the row before the last
$J_{D-3,i}$ are the KK momentum currents and the element
$J_{D-3,D-4}$ is the current for the $3d$ gauge field $\omega$.
Regular $4D$ solutions must have zero sources for this current,
otherwise $\omega$ will not be single valued.

\subsection{Nilpotency}

We now show that all the attractor flows are generated by the
nilpotent generators in the coset algebra. To get extremal black
hole solutions with a near horizon $AdS_2 \times S^2$, the
function $u$ must scale as $r^2$ as $r$ goes to zero while the
scalars go to a constant. This makes $S$ diverge as
$\frac{1}{r^2}$. The most natural way for $S$ to diverge as
$\tau^2$ for large $\tau$ is that $k$ is nilpotent, with
\begin{equation}
k^3=0 \ .
\end{equation}
This will be the crucial condition through
the whole paper.

\subsection{A Toy Example: Hyperk\"ahler Euclidean Metrics in
Four Dimensions}

Not every null geodesic corresponds to extremal black hole solutions.
Let's consider a simple example: hyperk\"ahler euclidean metrics in $4D$.

Although this example is not about a black hole, it is still quite instructive.
The $3d$ sigma model is $SL(2)/SO(1,1)$, i.e. $AdS_2$.
The coset representative is written as
\begin{equation}
M=\left(\begin{array}{cc} u^{1/2} & 0 \\ \frac{a}{u^{1/2}} & \frac{1}{u^{1/2}} \end{array} \right)
\end{equation}
and the symmetric invariant
\begin{equation}
S=\left(\begin{array}{cc} u - \frac{a^2}{u} & -\frac{a}{u} \\ -\frac{a}{u} & -\frac{1}{u} \end{array} \right)
\end{equation}

A geodesic is the exponential of a Lie algebra element in the
orthogonal to the stabilizer. The stabilizer $SO(1,1)$ is
generated by $\sigma^1$. A null geodesic is hence the exponential
of $k=\sigma^3 \pm i \sigma^2$. This is a nilpotent matrix,
$k^2=0$, hence $M= 1 + \tau k/2$. Take:
\begin{equation}
k=\sigma^3 + i \sigma^2=\left(\begin{array}{cc} 1 & 1 \\ -1 & -1
\end{array} \right)
\end{equation}
then
\begin{equation}
M=\left(\begin{array}{cc} 1+\tau/2 & \tau/2 \\ -\tau/2 & 1-\tau/2
\end{array} \right)
\end{equation}
and we can read off the geodesic solution from the invariant
\begin{equation}
S=M^T \left(\begin{array}{cc} 1 & 0 \\ 0 & -1
\end{array} \right) M=\left(\begin{array}{cc} 1+\tau & \tau \\ \tau & -1+\tau \end{array}
\right)
\end{equation}
Hence
\begin{equation}
u = \frac{1}{1-\tau},\qquad \qquad a = -\frac{\tau}{1-\tau}.
\end{equation}
Dualizing $a$, we get
\begin{equation}*d\omega = - \frac{da}{u^2} =
\frac{1}{r^2}dr \qquad \qquad u^{-1} = 1+\frac{1}{r}
\end{equation}

The 4D Euclidean metric is just the Taub-NUT metric. Notice that
the multi-centered Taub-NUT generalization of the metric is
obtained by replacing $\tau$ above with some harmonic function
$\sum_i \frac{q^i}{|x-x_i|}$. The sigma model equations of motion
are equivalent to the conservation of the current $J=S^{-1} \nabla
S$, and if $S$ is given as above with $\tau=\tau(\vec{x})$ then
the equation of motion are
\begin{equation}
\nabla^2 \tau(\vec{x})=0
\end{equation}

\subsection{Single-centered Black Holes in Pure Gravity}

\subsubsection{Constructing the flow generator $k$}

Now we look in detail at the single-centered black holes in pure
gravity. Notice that as $u$ goes to zero $S$ tends asymptotically
to a rank one matrix
\begin{equation}
S=- \frac{1}{u \sqrt{\det \rho}} \left(\begin{array}{ccc}\alpha_p \alpha_q & \alpha_p \alpha_0 &  \alpha_p \\
\alpha_0 \alpha_q &  \alpha_0 \alpha_0 & \alpha_0 \\
\alpha_q &
\alpha_0 & 1 \end{array} \right)
\end{equation}
hence the matrix $k^2$ should also have rank $1$. By inspection of
$S$ it is clear that a $k^2$ of rank higher than one gives a
geodesic for which the matrix elements of $\rho$ also diverge as
$\tau^2$ so that the scalar fields do not converge to fixed
attractor values.

Notice that if $k$ is nilpotent, then $S$ is a polynomial in
$\frac{1}{r}$, and the various scalars in the solution will all be
simple functions of $r$ for such extremal solutions!

The explicit form for $k$ in terms of the charges is then
straightforward to write. Consider the Jordan form of $k^2$: as it
is nilpotent, the eigenvalues are all zero. As it is rank one, it
has one single indecomposable block of size two:
$\left(\begin{array}{cc} 0 & 1 \\ 0 & 0 \end{array} \right)$. It
is written as
\begin{equation}
k^2 = -\eta v v^T f(P,Q)
\end{equation}with $v$ null in the metric
$\eta$, and $f(P,Q)$ any degree-two homogeneous function of the
charge $(P,Q)$. This form is chosen so that $v$ does not scale
with the charge $(P,Q)$.

Then $k$ must have a Jordan form with all eigenvalues
zero, one block of size $3$: $\left(\begin{array}{ccc} 0 & 1 &0 \\
0 & 0 & 1 \\ 0 & 0 & 0 \end{array} \right)$ and possibly some
other extra blocks of size two. Alternatively, there is a subspace
$V$ annihilated by $k$, a subspace $V'$ whose image under $k$ sits
in $V$ and has the same dimension as $V'$, and a single vector $w$
such that $kw \in V'$ and is non-zero.
\begin{equation}
kw \subset V', \qquad \qquad kV'\subset V, \qquad \qquad kV=0.
\end{equation}
From the symmetry of $\eta K$, it follows that the space $kV'$ is
made of null vectors only, and that $kw$ is orthogonal to it.
Because $\eta k^2 = -v v^T$, $(kw)^T \eta kw$ is negative. Because
of the signature of $\eta$ it is straightforward to see that $V'$ can
be of dimension at most one; hence there are no blocks of size $2$
in the Jordan form of $k$.

Taking this into account, the final form of $k$ is simply
\begin{equation}
k=\eta w v^T+\eta v w^T
\end{equation}
where $v$ and $w$ are two orthogonal $(D-2)$-dimensional vectors
with $v$ being null and $w$ having norm $-f(P,Q)$:
\begin{equation}\label{vwnorm}
w \eta v=0, \qquad v \eta v=0,\qquad w \eta w=-f(P,Q).
\end{equation}
Using the fact that in $k$, the first $D-4$ elements of the last
column $K_{i,D-2}$ are the magnetic charges, and the first $D-4$
elements of the row before the last $K_{D-3,i}$ are the electric
charges, and the element $K_{D-3,D-4}$ is the Taub-NUT charge,
which has to vanish, we have $2(D-4)+1=2D-7$ conditions. Together
with the three constraints coming from the norms and orthogonality
condition (\ref{vwnorm}), they can be used to solve for the $2D-4$ degrees of
freedom in $(v, w)$.

The full solution of $(v,w)$ requires one to solve some
degree-four equations, hence we'll leave it in a slightly implicit
form. Let $(p,q)$ be two $(D-4)$-dimensional vectors proportional to
the magnetic and electric charges, so that the magnetic charge and
the electric charge $(P,Q)$ of the $4D$ gauge fields are
\begin{equation}
P = \sqrt{p^2+p \cdot q}\,p \qquad \qquad Q=\sqrt{q^2+p \cdot
q}\,q
\end{equation}
And we choose $f(P,Q)$ to be
\begin{equation}
f(P,Q)=p\cdot q
\end{equation}
The solution of $v$ and $w$ written in terms of $(p,q)$ is
\begin{equation}
v =\frac{1}{\sqrt{p\cdot q}}\left(\begin{array}{c} q+p\\
-\sqrt{q^2+p\cdot q}\\
-\sqrt{p^2+p\cdot q}
\end{array} \right),\qquad\qquad
w
=\frac{\sqrt{ p \cdot q}}{2}\left(\begin{array}{c} q-p\\
-\sqrt{q^2+p\cdot q}\\
\sqrt{p^2+p\cdot q}
\end{array} \right).
\end{equation}
and $k$ can be written as
\begin{equation}
\label{PureGravityK}
k =\left(\begin{array}{ccc} q q^T- p p^T & -\sqrt{q^2+p \cdot q}\,
q&
\sqrt{p^2+p \cdot q}\, p \\
\sqrt{q^2+p \cdot q} \,q^T & -(q^2+p \cdot q) & 0 \\
-\sqrt{p^2+p \cdot q}\, p^T & 0 & p^2+p \cdot q \end{array}
\right)
\end{equation}

\subsubsection{Full flow}

First, for the full flow starting from $M_0=1$, the scalars for
the attractor solution generated by this $k$ can be read off from
$S(\tau)=e^{k\tau}$, by comparing with the form of $S$ in terms of
the 4D fields:
\begin{equation}
u^{-2} = [1+(p^2+p \cdot q)(\tau + \frac{p \cdot
q}{2}\tau^2)][1+(q^2+p \cdot q)(\tau + \frac{p \cdot q}{2}\tau^2)]
\end{equation}
and
\begin{equation}
\rho = 1 + \frac{ (q q^T- p p^T)\tau + [(p^2+p \cdot q) q q^T -
\frac{p \cdot q}{2} (p+q)(p+q)^T]\tau^2}{1+(p^2+p \cdot q)(\tau +
\frac{p \cdot q}{2}\tau^2)}
\end{equation}
Notice that as $\tau \rightarrow \infty$, $\tau^{-2} e^{-2U}$  has
the correct limit $\frac{P\cdot Q}{2}$, which is the entropy where
$P$ and $Q$ are the physical electric and magnetic charges.

To generalize to arbitrary asymptotic moduli,
$M(\tau)=e^{g/2}e^{k\tau/2}$, and the flow of $S$ is
(\ref{Stauarbitrymoduli}), which can be written as
$S(\tau)=e^{K(\tau)}S_0$, where $K(\tau)$ is a matrix function.
From now on, we use lower case $k$ to denote the coset algebra
that generates the attractor flow, and capital $K$ to denote the
function which we exponentiate directly to produce the solution.

We will choose $K(\tau)$ to have the same properties as the
generator $k$:
\begin{equation}
K^3(\tau)=0 \qquad \textrm{and} \qquad K^2(\tau) \,\,\textrm{rank
one}
\end{equation}
The equations of motion $\nabla \cdot (S^{-1} \nabla S)=0$ then
simplify considerably with this ansatz. If one further requires
that the subspace image of $K^2(\tau)$ remains constant
everywhere, such that
\begin{equation}
K^2(\tau)\nabla K(\tau) = \nabla K(\tau) K^2(\tau) =0
\end{equation}then the current reduces to
\begin{equation}
J =S^{-1} \nabla S=S_0\left(\nabla K(\tau) + \frac{1}{2}[\nabla
K(\tau),K(\tau)]\right)S_0
\end{equation} and the equations of
motion are
\begin{equation}
\nabla^2 K(\tau) + \frac{1}{2}[\nabla^2 K(\tau),K(\tau)]=0
\end{equation}
which is solved by a harmonic $K(\tau)$.

It might appear hard to find a $K(\tau)$ that is harmonic and
satisfies all the required constraints. However, by remembering
that the constraints dictate $K(\tau)$ to have the form:
\begin{equation}
K(\tau)= \eta V W^T  + \eta W V^T
\end{equation}
with $V$ being null and doesn't scale with the charge $(P,Q)$, and
$W$ orthogonal to $V$ everywhere, one can simply pick a constant
null vector $V=v'$ and a harmonic vector $W(\tau)$ everywhere
orthogonal to $v'$:
\begin{equation}
W(\tau)=w'\tau+m \qquad \textrm{with} \qquad v'\cdot W(\tau)=0
\end{equation}
Here $m$ is a $(D-2)$-vector and contains the information of
asymptotic moduli. Thus an appropriate $K(\tau)$ is built:
\begin{equation}
K(\tau)=k'\tau+g
\end{equation}
where
\begin{equation}
k'=\eta v'w'^T+\eta w'v'^T \qquad \qquad g=\eta v'm^T+\eta mv'^T
\end{equation}

Now we need to solve for $(v',w')$ for the same charge $(P,Q)$ but
in the presence of $m$. The form of $g$ guaranteed that
\begin{equation}
[k',g]=0
\end{equation}
where we have used the fact that $v'$ is null and $w'$ is orthogonal
to $v'$. Therefore, shifting the starting point of moduli does not
change the current as a function of $(v,w)$:
\begin{equation}
J(v',w')=S_0\left(\frac{k'}{r^2}\right)S_0=S_0\left(\frac{\eta
v'(w')^T+\eta w'(v')^T}{r^2}\hat{\vec{r}}\right)S_0
\end{equation}
Thus, the solution of $(v',w')$ in terms of charges solved from
the current does not change as we vary the starting point of the
flow, i.e. they do not depend on the asymptotic moduli:
\begin{equation}
v'(Q)=v(Q) \qquad \qquad w'(Q)=w(Q)
\end{equation}

In summary, the flow with arbitrary starting point is simply
generated by
\begin{equation}
K(\tau)=\eta vW(\tau)^T+\eta W(\tau)v^T  \qquad \textrm{with}
\qquad W=w\tau+m
\end{equation}
where $(v,w)$ only depend on the charges $(P,Q)$ and $m$ gives the
asymptotic moduli.

\subsubsection{Example: 5D pure gravity compactified on a circle.}
Consider for example the case of extremal black holes in $D=5$
pure gravity compactified on a circle. The $3d$ sigma model is
$SL(3)/SO(1,2)$. The symmetric invariant is
\begin{equation}
S_{gr}=\left(\begin{array}{ccc} \rho - \frac{1}{u \sqrt{ \rho}}
\alpha_1 \alpha_1 & \rho A_0 - \frac{1}{u \sqrt{ \rho}} \alpha_1
\alpha_0 & -\frac{1}{u \sqrt{\rho}} \alpha_1 \\ \rho A_0 -
\frac{1}{u \sqrt{ \rho}} \alpha_1 \alpha_0 & \rho (A_0)^2-
\frac{u}{ \sqrt{\rho}}- \frac{1}{u \sqrt{ \rho}} \alpha_0 \alpha_0
& -\frac{1}{u \sqrt{\rho}} \alpha_0 \\ -\frac{1}{u \sqrt{ \rho}}
\alpha_1 & -\frac{1}{u \sqrt{\rho}} \alpha_0 & -\frac{1}{u
\sqrt{\rho}} \end{array} \right)
\end{equation}
Then we can calculate $S = S_0 e^{ k\tau}$
using (\ref{PureGravityK}) and compare the result to $S_{gr}$ above to solve for all the scalars.
We find that
\begin{eqnarray}
e^{-2U} &=& \sqrt{[1+(q^2+pq)(\tau+\frac{pq}{2}
\tau^2)][1+(p^2+pq)(\tau+\frac{pq}{2}\tau^2)]}\\
\rho &=& \frac{1+(q^2+pq)(\tau+\frac{pq}{2}
\tau^2)}{1+(p^2+pq)(\tau+\frac{pq}{2}\tau^2)}
\end{eqnarray}
when starting from the identity. If we allow arbitrary $g$ the
flow is too complicated to write explicitly here, but the
attractor value of $\rho$ is the same: $q/p$. 

\subsection{Multi-centered Solutions in Pure Gravity}

In the context of pure gravity compactified on a torus, we can
also give some examples of multi-centered solutions in the same
spirit as the ones for BPS solutions in $N=2$ supergravity, though
some important features of the latter are not present here.

We are interested in solutions given in terms of harmonic
functions which can generalize the single-centered extremal
solutions presented above. Similar to the single-centered case, we
exponentiate a matrix function $K(\vec{x})$:
\begin{equation}
S(\vec{x})=e^{K(\vec{x})}S_0
\end{equation}
We will choose $K(\vec{x})$ to have the same properties of the
generator $k$:
\begin{equation}
K^3(\vec{x})=0 \qquad \textrm{and} \qquad K^2(\vec{x})
\,\,\textrm{rank one}
\end{equation}

Using a similar argument to the single-centered flow, we require
that the subspace image of $K^2(\vec{x})$ remains constant
everywhere, such that $K^2(\vec{x})\nabla K(\vec{x}) = \nabla
K(\vec{x}) K^2(\vec{x}) =0$, then the equations of motion are
\begin{equation}
\nabla^2 K(\vec{x}) +\frac{1}{2} [\nabla^2
K(\vec{x}),K(\vec{x})]=0
\end{equation}
which is solved by a harmonic $K(\vec{x})$.

A multi-centered $K(\vec{x})$ that is harmonic and satisfies all
the required constraints can then be built in the same way as the
single-centered one:
\begin{equation}
K(\vec{x})=\eta v W(\vec{x})^T+\eta W(\vec{x})v^T
\end{equation}
where $v$ is the same constant null vector as in $k$, and
$W(\vec{x})$ is the multi-centered harmonic function:
\begin{equation}
W(\vec{x})=\sum_i\frac{w_i}{|\vec{x}-\vec{x}_i|}+m
\end{equation}
where $\vec{x}_i$ is the position of the $i$th center, and $w_i$ is
determined by the charges at the $i$th center, and $m$ is related to
the moduli at infinity.   Requiring $W(\vec{x})$ to be
orthogonal to $v$ everywhere gives the following constraints on
the $\{w_i,m\}$: First, taking $\vec{x}$ to infinity, it gives
\begin{equation}
v \cdot m=0
\end{equation}
Second, $w_i$ are orthogonal to $v$:
\begin{equation}
v\cdot w_i=0
\end{equation}
In addition to the constraint from the zero Taub-NUT charge
condition $-bz-cy=0$, this makes the space of each possible $w_i$
only $(D-4)$-dimensional. That is, though $W$ is a $(D-2)$-vector,
one has only $(D-4)$ independent harmonic functions to work with,
because of the orthogonality to $v$ and the requirement of no
timelike NUT charges. This makes the solution relatively boring.

The multi-centered solution in pure gravity does not have the
characteristic features of the typical BPS multi-centered solution
in $\mathcal{N}=2$ supergravity, where many centers with
relatively non-local charges form bound states which carry
intrinsic angular momentum.

The basic reason is that when the ansatz $K(\vec{x})= \eta v
W^T(\vec{x})  + \eta W(\vec{x}) v^T$ is used, the second term of
the conserved currents $J = S_0(\nabla K +\frac{1}{2} [\nabla
K,K])S_0$ drops out. The first result is that the charges of the
various centers in the solution can be read off directly from $(v,
w_i)$, and they do not depend on the positions, charges of the
other centers. Thus, there is no constraint on the position of
each center as in the $\mathcal{N}=2$ BPS multi-centered solution;
centers can be moved around freely.

Moreover, the condition of no timelike Taub-NUT charge is a linear
constraint on the charges at each center which results in a static
$4D$ solution, as
$*d\omega=0$
 leads to
$\omega=0$.
Therefore, no angular momentum is present.

\section{Attractor Flows in $G_{2(2)}/(SL(2,\mathbb{R})\times SL(2,\mathbb{R}))$}

We now tackle a more complicated subject: ${\cal N}=2$
supergravity coupled to one vector multiplet. First, we reduce the
theory down to three dimensions and derive the metric for the
resulting moduli space, which is the coset
$G_{2(2)}/(SL(2,\mathbb{R})\times SL(2,\mathbb{R}))$.\footnote{Other work on this coset space has appeared recently, including \cite{Bouchareb:2007ax, Clement:2007qy, Gunaydin:2007qq}.} We then
discuss the Cartan and Iwasawa decompositions of the group
$G_{2(2)}$, which we use to construct the coset algebra and
translate the flow of coset representative into the flow of the
moduli fields, respectively. We then specify the representation of
$G_{2(2)}$ we will be working with, and describe the form of
attractor flow generators in this representation.

\subsection{The moduli space $\mathcal{M}_{3D}$}

The $3d$ moduli space for ${\cal N}=2, d=4$ supergravity coupled to $n_V$ vector multiplets is well-studied, for example in \cite{de Wit:1992wf,Ceresole:1995ca,Pioline:2006ni}. Some of the main results are compiled in the Appendix. Here we briefly review the essential points.

The bosonic part of the action is:
\begin{equation}
S=-\frac{1}{16\pi}\int d^4x
\sqrt{g^{(4)}}\left[R-2g_{i\bar{j}} dz^i\wedge \ast_4
d\bar{z}^{\bar{j}}-F^I\wedge G_I\right]
\end{equation}
where $I=0,1...n_V$, and $G_I =  (Re{\cal N})_{IJ} F^J+(Im{\cal
N})_{IJ} \ast F^J$.
For a model endowed with a prepotential $F(X)$,
\begin{equation}
\mathcal{N}_{IJ}=F_{IJ}+2i\frac{(\textrm{Im} F\cdot
X)_I(\textrm{Im} F \cdot X)_J}{X \cdot \textrm{Im} F \cdot X}
\end{equation}
where $F_{IJ}=\partial_I \partial_J F(X)$. We reduce to three
dimensions, dualizing the vectors ($\omega$,$\mathbf{A}^I$) to the
scalars ($\sigma$,$B_I$), and renaming $A_0^I$ as $A^I$. The
metric of $\mathcal{M}^{*}_{3D}$ is then
\begin{eqnarray}
\label{G22Metric} ds^2&=&\textbf{d}U \cdot \textbf{d}U
+\frac{1}{4}e^{-4U}(\textbf{d}\sigma+A^I \textbf{d}B_I-B_I
\textbf{d}A^I)\cdot (\textbf{d}\sigma+A^I \textbf{d}B_I-B_I
\textbf{d}A^I)+g_{i\bar{j}}(z,\bar{z})\textbf{d}z^i \cdot
\textbf{d}\bar{z}^{\bar{j}}\non\\
&&+\frac{1}{2}e^{-2U}[(Im\mathcal{N}^{-1})^{IJ}(\textbf{d}B_I+\mathcal{N}_{IK}\textbf{d}A^K)\cdot
(\textbf{d}B_J+\overline{\mathcal{N}}_{JL}\textbf{d}A^L) ]
\end{eqnarray}
It is a para-quaternionic-K\"ahler manifold. Since the holonomy is
reduced from $SO(4n_V+4))$ to $Sp(2,\mathbb{R})\times
Sp(2n_V+2,\mathbb{R})$, the vielbein has two indices $(\alpha,A)$
transforming under $Sp(2,\mathbb{R})$ and $Sp(2n_V+2,\mathbb{R})$,
respectively. The para-quaternionic vielbein is the analytical
continuation of the quaternionic vielbein computed in
\cite{Ferrara:1989ik}.  The explicit form is given in the
appendix.

For $n_V=1$, $X^I=(X^0,X^1)$. For our purpose, we choose the
prepotential
\begin{equation}
F(X)=-\frac{(X^1)^3}{X^0}
\end{equation}
The metric of $\mathcal{M}^{*}_{3D}$ with one-modulus is
(\ref{G22Metric}) with $g_{z\bar{z}}=\frac{3}{4y^2}$ and
$\mathcal{N}$ and $(Im\mathcal{N})^{-1}$ being
\begin{displaymath}
\mathcal{N}=\left(%
\begin{array}{cc}
-(2x-iy)(x+iy)^2  &3x(x+iy) \\
   3x(x+iy)&-3(2x+iy)\\
\end{array}%
\right)\qquad Im\mathcal{N}^{-1}=-\frac{1}{y^3}\left(%
\begin{array}{cc}
1  &x \\
   x&3x^2+y^2\\
\end{array}%
\right)
\end{displaymath}

The isometries of the $\mathcal{M}^{*}_{3D}$ descend from the
symmetries of the 4D system. The gauge symmetries in 4D give
shifting isometries of $\mathcal{M}^{*}_{3D}$, whose associated conserved charges are:
\begin{equation}
\label{ConservedChargesCurrents}
q_I d\tau =
J_{A^I}=P_{A^{I}}-B_{I}P_{\sigma}, \qquad \qquad p^I
d\tau=J_{B_I}= P_{B_{I}}+A^{I}P_{\sigma}, \qquad \qquad k
d\tau=J_{\sigma}=P_{\sigma}
\end{equation}
where the
momenta $\{P_{\sigma},P_{A^I},P_{B_I}\}$ are
\begin{eqnarray}
\label{SigmaConservedCurrent}
P_{\sigma}&=&\frac{1}{2}e^{-4U}(\textbf{d}\sigma+A^I
\textbf{d}B_I-B_I
\textbf{d}A^I)\\
P_{A^{I}}&=& e^{-2U}[(Im\mathcal{N})_{IJ}
\textbf{d}A^J+(Re\mathcal{N})_{IJ}(Im\mathcal{N}^{-1})^{JK}(\textbf{d}B_K+(Re\mathcal{N})_{KL}\textbf{d}A^L)]-B_IP_{\sigma}\\
P_{B_{I}}&=&
e^{-2U}[(Im\mathcal{N}^{-1})^{IJ}(\textbf{d}B_J+(Re\mathcal{N})_{JK}\textbf{d}A^K)]+A^IP_{\sigma}
\end{eqnarray}
Here $\tau$ is the affine parameter defined as $d\tau \equiv
-\boldsymbol{\ast}_{3}\sin{\theta}d\theta d\phi$.
$(p^0,p^1,q_1,q_0)$ are the D6-D4-D2-D0 charges, and $k$ the
Taub-NUT charge. A non-zero $k$ gives rise to closed time-like
curves, so we will set $k=0$ from now on.

Note that
the time translational invariance in 4D gives rise to the
conserved current
\begin{equation}
J_{U}=P_{U}+2\sigma J_{\sigma}+A^IJ_{A^I}+B^IJ_{B^I}
\end{equation}
where $P_U=2dU$. The corresponding conserved charge is the ADM
mass: $2M_{ADM}d\tau=J_{U}$.

\subsection{Extracting the Coordinates from the Coset Elements}

\begin{figure}[htbp]
  \centering
  \includegraphics{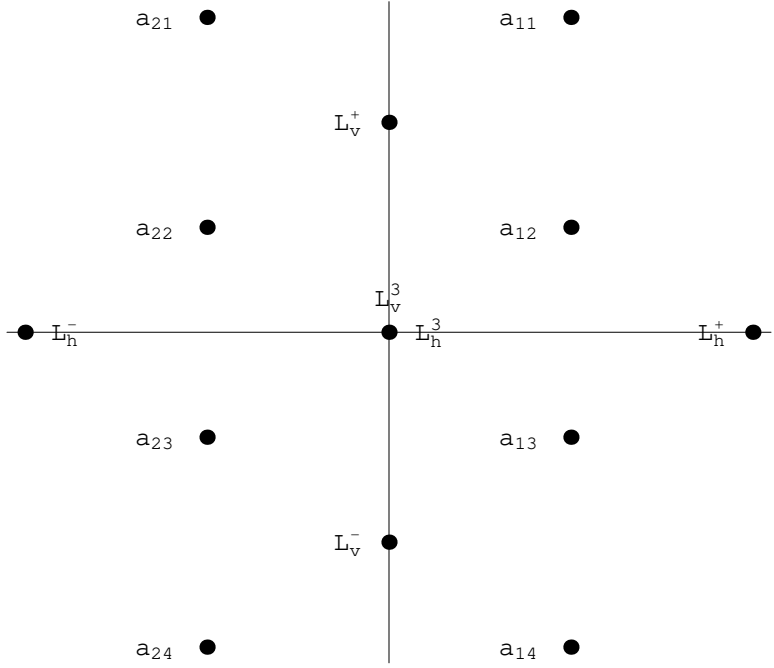}
  \caption{Root Diagram of Cartan Decomposition of $G_{2(2)}$}
  \label{fig:RootCartan}
\end{figure}

 The metric (\ref{G22Metric}) for the case $n_V=1$ describes an eight-dimensional
manifold with coordinates $\phi^n=\{u,x,y,\sigma,A^0, A^1,B_1,
B_0\}$. This manifold is the coset space $G_{2(2)}/(SL(2,\mathbb{R})\times
SL(2,\mathbb{R}))$. The root diagram for the Cartan decomposition
of $G_{2(2)}$ is shown in Figure \ref{fig:RootCartan}.
The six roots that lie on the horizontal and vertical axes
$\{L^{\pm}_h, L^3_h, L^{\pm}_v, L^3_v\}$ are the six non-compact
generators of the subgroup $\textbf{H}=SL(2,\mathbb{R})_{h}\times
SL(2,\mathbb{R})_{v}$:
\begin{equation}
[L^3_{h/v},L^{\pm}_{h/v}]=\mp L^{\pm}_{h/v}, \qquad
[L^{+}_{h/v},L^{-}_{h/v}]=2L^3_{h/v}
\end{equation} and the two vertical columns of eight roots $\{a_{\alpha A}\}$
are the basis of the subspace \textbf{K}. $\{a_{1A},a_{2A}\}$ for
each $A$ is a spin-$1/2$ doublet under the horizontal
$SL(2,\mathbb{R})$:
\begin{displaymath}
[%
L^3_{h},\left(\begin{array}{c}
  a_{1 A} \\
  a_{2 A}
\end{array}%
\right) ]=\left(\begin{array}{c}
 -\frac{1}{2} a_{1 A} \\
 \frac{1}{2}a_{2 A}
\end{array}%
\right) \qquad [%
L^{+}_{h},\left(\begin{array}{c}
   a_{1 A} \\
  a_{2 A}
\end{array}%
\right) ]=\left(\begin{array}{c}
 0 \\
  a_{1 A}
\end{array}%
\right)\qquad  [%
L^{-}_{h},\left(\begin{array}{c}
   a_{1 A} \\
  a_{2 A}
\end{array}%
\right) ]=\left(\begin{array}{c}
  -a_{2 A} \\
  0
\end{array}%
\right)
\end{displaymath}
And $\{a_{\alpha 1},a_{\alpha 2}, a_{\alpha 3},a_{\alpha 4}\}$ for
each $\alpha$ span a spin-$3/2$ representation of the vertical
$SL(2,\mathbb{R})$:
\begin{displaymath}
[%
L^3_{v},\left(\begin{array}{c}
  a_{\alpha 1} \\
  a_{\alpha 2} \\
  a_{\alpha 3}\\
  a_{\alpha 4}
\end{array}%
\right) ]=\left(\begin{array}{c}
 -\frac{3}{2} a_{\alpha 1} \\
 -\frac{1}{2}a_{\alpha 2} \\
 \frac{1}{2} a_{\alpha 3}\\
 \frac{3}{2}a_{\alpha 4}
\end{array}%
\right) \qquad [%
L^{+}_{v},\left(\begin{array}{c}
  a_{\alpha 1} \\
  a_{\alpha 2} \\
  a_{\alpha 3}\\
  a_{\alpha 4}
\end{array}%
\right) ]=\left(\begin{array}{c}
 0 \\
 3a_{\alpha 1} \\
 2a_{\alpha 2}\\
 a_{\alpha 3}
\end{array}%
\right)\qquad  [%
L^{-}_{v},\left(\begin{array}{c}
  a_{\alpha 1} \\
  a_{\alpha 2} \\
  a_{\alpha 3}\\
  a_{\alpha 4}
\end{array}%
\right) ]=\left(\begin{array}{c}
 -a_{\alpha 2} \\
 -2a_{\alpha 3} \\
 -3a_{\alpha 4}\\
 0
\end{array}%
\right)
\end{displaymath}
All the commutators can be easily read off from the Root diagram
(\ref{fig:RootCartan}), we will only write down the following ones
which will be useful later.
\begin{equation}\label{aacommu}
[a_{11},a_{14}]=-\frac{1}{3}[a_{12},a_{13}]=-4L^{+}_{h} \qquad
[a_{21},a_{24}]=-\frac{1}{3}[a_{22},a_{23}]=-4L^{-}_{h}
\end{equation}

\begin{figure}[htbp]
  \centering
  \includegraphics{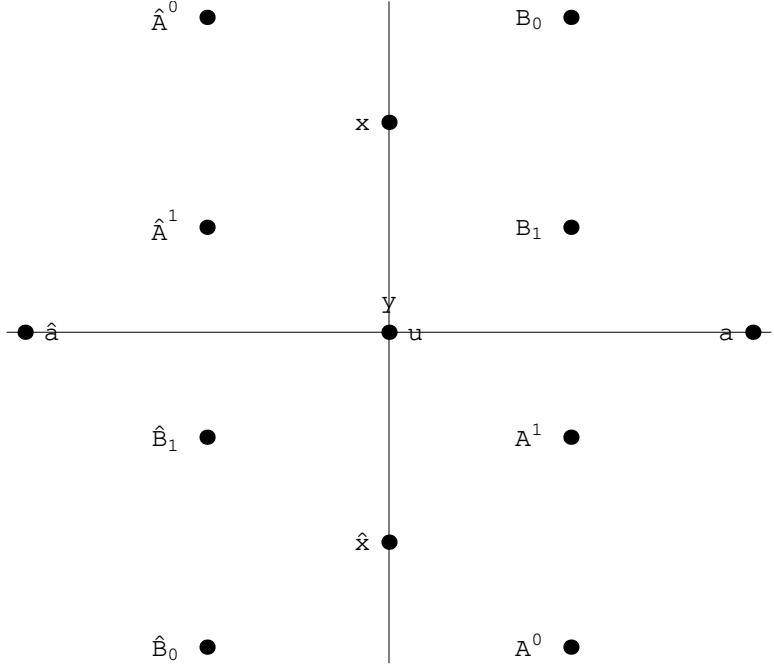}
  \caption{Root Diagram of Isometry of $M_{3D}=G_{2(2)}/(SL(2,\mathbb{R})\times SL(2,\mathbb{R}))$. $\{\mathbf{u},\mathbf{y},\mathbf{x},
\bm{\sigma},\mathbf{A}^{0},\mathbf{A}^{1},\mathbf{B}_{1},\mathbf{B}_{0}\}$
generates the solvable subgroup.}
  \label{fig:RootSolv}
\end{figure}

Being semisimple, the algebra of $G_{2(2)}$ has the Iwasawa
decomposition
$\mathbf{g}=\mathbf{h}\oplus\mathbf{a}\oplus\mathbf{n}$, where
$\mathbf{a}$ is the maximal abelian subspace of $\mathbf{k}$, and
$\mathbf{n}$ is the nilpotent subspace of the positive root space
$\Sigma^{+}$ of $\mathbf{a}$. In Figure \ref{fig:RootSolv}, we
show the Iwasawa decomposition of $G_{2(2)}$.
The two Cartan generators in $\mathbf{a}$ are $\{\mathbf{u},
\mathbf{y}\}$, and $\{\mathbf{x},
\bm{\sigma},\mathbf{A}^{0},\mathbf{A}^{1},\mathbf{B}_{1},\mathbf{B}_{0}\}$
span a nilpotent subspace $\mathbf{n}$: $n^7=0$ for $n
\in\mathbf{n}$. $\mathbf{a}$ and $\mathbf{n}$ together generate
the solvable subgroup $Solv$ of $\mathbf{G}$, which act
transitively on $\mathcal{M}_{3D}=G_{2(2)}/SL(2,\mathbb{R})\times
SL(2,\mathbb{R})$. In particular, $\mathbf{y}$ generates the
rescaling of $y$, and $\{\mathbf{u},\mathbf{x},
\bm{\sigma},\mathbf{A}^{0},\mathbf{A}^{1},\mathbf{B}_{1},\mathbf{B}_{0}\}$
generates the translation of $\{U,x,\sigma,A^0,A^1,B_1,B_0\}$. The
moduli space $\mathcal{M}_{3D}$ can be parameterized by the
solvable elements:
\begin{equation}
\Sigma(\phi)=e^{U\mathbf{u}+(\ln{y})\mathbf{y}}e^{x\mathbf{x}+A^I\mathbf{A}^I+B_I\mathbf{B_I}+\sigma\bm{\sigma}}
\end{equation}
The origin of the moduli space
\begin{equation}\label{origin}
a=A^0=A^1=B_1=B_0=0 \qquad x=0\qquad y=u=1
\end{equation}correspond to $\Sigma(\phi)=1$.

In Fig \ref{fig:RootSolv}, the isometries are plotted according to
their eigenvalues under the two Cartan generators $\mathbf{u}$ and
$\mathbf{y}$ \cite{de Wit:1992wf}. $\{\mathbf{u},\mathbf{y}\}$ are
related to $a_{\alpha A}$ by\footnote{The matrix representation of
$\mathbf{u}$ and $\mathbf{y}$ are
\begin{equation}\mathbf{u}=Diag[0,\frac{1}{2},-\frac{1}{2},0,-\frac{1}{2},\frac{1}{2},0]\qquad \mathbf{y}=Diag[1,-\frac{1}{2},-\frac{1}{2},-1,\frac{1}{2},\frac{1}{2},0]
\end{equation}}:
\begin{equation}
\mathbf{u}=-\frac{1}{8}[(a_{11}+a_{24})-(a_{13}+a_{22})] \qquad
\qquad \mathbf{y}=\frac{1}{8}[3(a_{11}+a_{24})+(a_{13}+a_{22})]
\end{equation}
The three generators
$\{\bm{\sigma},\mathbf{u},\hat{\bm{\sigma}}\}$ on the horizontal
axis and $\{\mathbf{x},\mathbf{y},\mathbf{\hat{x}}\}$ on the
vertical axis form the horizontal and vertical $SL(2,\mathbb{R})$,
respectively. The vertical $SL(2,\mathbb{R})$ generate the duality
invariance. Denote the two vertical columns of eight isometries as
\begin{displaymath}
\left(%
\begin{array}{cc}
 \bm{\xi}_{21}& \bm{\xi}_{11}\\
  \bm{\xi}_{22}& \bm{\xi}_{12}\\
 \bm{\xi}_{23}& \bm{\xi}_{13}\\
  \bm{\xi}_{24}& \bm{\xi}_{14}
\end{array}%
\right) \equiv\left(%
\begin{array}{cc}
-\mathbf{\hat{A}}^0& \mathbf{B}_0\\
 -3\mathbf{\hat{A}}^1&3\mathbf{B}_1\\
 \mathbf{\hat{B}}_1&-\mathbf{A}^1\\
 -\mathbf{\hat{B}}_0&\mathbf{A}^0
\end{array}%
\right)
\end{displaymath}
$\{\bm{\xi}_{1 A},\bm{\xi}_{2 A}\}$ for each $A$ span a spin-$1/2$
representation of the horizontal $SL(2,\mathbb{R})$, and
$\{\bm{\xi}_{\alpha 1},\bm{\xi}_{\alpha 2}, \bm{\xi}_{\alpha
3},\bm{\xi}_{\alpha 4}\}$ for each $\alpha$ span a spin-$3/2$
representation of the vertical $SL(2,\mathbb{R})$.

Parameterizing the coset representative $M$ as the solvable
elements, the symmetric matrix $S$ can be written in terms of the
eight coordinates $\phi^n$, from the solvable elements $\Sigma$
\begin{equation}
S(\phi)=\Sigma(\phi)S_0\Sigma(\phi)^T
\end{equation}
The coordinates are read off from the symmetric matrix $S$.

\subsection{Nilpotency of the Attractor Flow Generator $k$.}

The near-horizon geometry of the $4D$ attractor is $AdS_2 \times
S^2$, i.e.
\begin{equation}
e^{-U}\rightarrow \sqrt{V_{BH}}|_{*} \tau  \qquad \textrm{as}
\qquad \tau \rightarrow \infty
\end{equation}
In terms of the variable $u\equiv e^{2U}$
\begin{equation}
u \rightarrow \frac{1}{V_{BH}}\tau^{-2} \qquad \textrm{as} \qquad
\tau \rightarrow \infty
\end{equation}
The solvable element is
\begin{equation}
M=e^{U\mathbf{u}+\dots}\sim u^{\frac{1}{2}\mathbf{u}}
\end{equation}
As the flow goes to the near-horizon, $u \rightarrow 0$
\begin{equation}
M(\tau)\sim u^{-\ell/2}\sim \tau^{\ell}
\end{equation}
where $-\ell$ is the lowest eigenvalue of $\mathbf{u}$. That is,
$M(\tau)$ is a polynomial function of $\tau$.

On the other hand, since the geodesic flow is generated by $k$ via
\begin{equation}
M(\tau)=M(0)e^{k \tau/2}
\end{equation}
i.e., $M(\tau)$ is an exponential function of $\tau$. To reconcile
the two statements, $k$ must be nilpotent:
\begin{equation}
k^{\ell+1}=0
\end{equation}
That is, the element in $\mathbf{k}$ that generates the attractor
flow is nilpotent. Moreover, by looking at the weights of the
fundamental representation of $G_2$, we see that $\ell=2$
\begin{equation}
k_{*}^{3}=0
\end{equation}

Moreover, the nilpotency of the attractor flow generators
guarantees that it is null:
\begin{equation}
k^3=0 \qquad \Longrightarrow \qquad (k^2)^2=0 \Longrightarrow
Tr(k^2)=0
\end{equation}
which means that $k$ is null.

\subsection{Properties of Attractor Flow}

The scalar moduli space is parameterized by a symmetric $7 \times
7$ matrix $S$ which sits in $G_{2(2)}$, i.e. preserves a
non-degenerate three form $w_{ijk}$ such that
$\eta_{is}=w_{ijk}w_{stu}w_{mno} \epsilon^{jktumno}$ is a metric
with signature $(4,3)$ normalized so that $\eta^2=1$. To
facilitate the comparison with the pure $5D$ gravity case we
decompose $7$ as $3 \oplus \bar 3 + 1$ of $SL(3)$ and pick as
non-zero components of $w$ the $3 \wedge 3 \wedge 3$, $\bar 3
\wedge \bar 3 \wedge \bar 3$ and $3 \otimes \bar 3 \otimes 1$ as
\begin{equation}
w=dx_1\wedge dx_2\wedge dx_3 + dy^1\wedge dy^2\wedge dy^3 -
\frac{1}{\sqrt{2}} dx_a \wedge dy^a \wedge dz
\end{equation}The
resulting expression for $\eta$ is
\begin{equation}
\eta= dx_ady^a - dz^2
\end{equation}

We know that $k$ must be an element of $G_{2(2)}$, hence also of
$SO(4,3)$. As in the pure gravity case, we choose the
representation such that
\begin{equation}
S_0 k S_0 = k^T \qquad \qquad S_0 k^2 S_0 = (k^2)^T
\end{equation}
In this base a $G_{2(2)}$ Lie algebra element is given as
\begin{equation}
\label{kBasis} k = \left(\begin{array}{ccc} A_{i1}^{j1} &
\epsilon_{i1 j2 k} v^k & \sqrt{2} w_{i1} \\ \epsilon^{i2 j1 k} w_k
& -A_{j2}^{i2} & -\sqrt{2} v^{i2} \\ -\sqrt{2} v^{j1} & \sqrt{2}
w_{j2} & 0
\end{array} \right)
\end{equation}
Here $A$ is a traceless $3\times 3$ matrix. $S$ is a symmetric
element in $G_{2(2)}$ with signature $\{1,-1,-1,1,-1,-1,1\}$, i.e.
$S=MS_0M^T$ with
\begin{equation}
S_{0} = \left(\begin{array}{ccc} \eta_1 & 0 & 0 \\ 0 & \eta_1 & 0
\\ 0 & 0 & 1 \end{array} \right)=Diag(1,-1,-1,1,-1,-1,1)
\end{equation}
where $\eta_1$ is the one for pure gravity.

If the gauge field is turned off, then $S$ is block diagonal
\begin{equation}
S|_{F=0} = \left(\begin{array}{ccc} S_{gr} & 0 & 0 \\ 0 &
S_{gr}^{-1} & 0 \\ 0 & 0 & 1 \end{array} \right)
\end{equation}
where $S_{gr}$ is the same as the one for pure $5D$ gravity.
Turning on a non-zero $5D$ vector field corresponds to a more
general $S$:
\begin{equation}
S=e^{k_1^T} (S|_{F=0})e^{k_1}
\end{equation}
with $k_1$ a $G_{2(2)}$ Lie algebra matrix with $w_1$ equal to the
fifth component of the gauge field, $v^2$ equal to the time
component of the gauge field and $w_3$ equal to the scalar dual to
the three-dimensional part of the gauge field.

In this representation, $(x,y)$ can be extracted from symmetric
matrix $S$ via:
\begin{equation}\label{xyfromS}
x(\tau)=-\frac{S_{35}(\tau)}{S_{33}(\tau)} \qquad
y^2=\frac{S_{33}(\tau)S_{55}(\tau)-S_{35}(\tau)^2}{S_{33}(\tau)^2}
\end{equation}
And $u$ via:
\begin{equation}\label{ufromS}
u=\frac{1}{\sqrt{S_{33}(\tau)S_{55}(\tau)-S_{35}(\tau)^2}}
\end{equation}

The $4D$ gauge currents sit in $J=S^{-1} \nabla S$, where
$J_{12}(J_{31})$ is again the electric(magnetic) current for the
$KK$ photon, $J_{32}$ the timelike NUT current, and
$J_{72}(J_{51})$ the electric(magnetic) current for the reduction
of the $5D$ gauge field.
\begin{equation}
\label{ExtractCharges}
J_{32}= -2 J_{\sigma} \qquad J_{12}= \sqrt{2}J_{A^0} \qquad
J_{72}= \frac{2}{3}J_{A^1} \qquad J_{51}=-\sqrt{2} J_{B_1} \qquad
J_{31}=\sqrt{2} J_{B_0}
\end{equation}Moreover,
\begin{equation}
J_{22}-J_{33}=2J_{U}
\end{equation}
We use $\mathbf{Q}$ to denote the charge matrix, where it relates
to the D-brane charge $\{p^0,p^1,q_1,q_0\}$ and the vanishing NUT
charge $k$ by
\begin{equation}\label{QDbranes}
(\mathbf{Q}_{31},\mathbf{Q}_{51},\mathbf{Q}_{72},\mathbf{Q}_{12})=(\sqrt{2}p^0,-\sqrt{2}p^1,\frac{2}{3}q_1,\sqrt{2}q_0)
\qquad \mathbf{Q}_{32}=-2k=0
\end{equation}

Since $k$ is nilpotent: $k^3=0$,
\begin{equation}
S=e^{k\tau}S_0 =(1+k\tau+\frac{1}{2} k^2\tau^2)S_0
\end{equation}
The $AdS_2\times S^2$ near-horizon geometry of the $4D$ attractor
dictates $u=\frac{1}{V_{BH}|_{*}} \tau^{-2}$ as $\tau\rightarrow
\infty$. Therefore, the flow generator $k$ can be obtained by
\begin{equation}
k^2=2 V_{BH}|_{*} (u S |_{u \rightarrow 0})S_0
\end{equation}
Computing $k^2$ using $S$ constructed from the solvable elements
$\Sigma(\phi)$ shows that $k^2$ is of rank two, its Jordan form
has two blocks of size $3$. It can be written as
\begin{equation}\label{kSquaredNullVectors}
k^2=\sum_{a,b=1,2} v_a v_b^T c_{ab}S_0
\end{equation}
with $v_a$ null and orthogonal to each other: $v_a \cdot v_b\equiv
v^T_aS_0v_b=0$, and $c_{ab}$ depends on the particular choice of
$k$. Thus $k$ can be expressed as:
\begin{equation}
\label{kInTermsOfvw} k=\sum_{a=1,2} (v_a w_a^T+ w_a v_a^T)S_0
\end{equation}
where each $w_a$ is orthogonal to both $v_a$: $w_a \cdot v_b=0$,
and $w_{a}$ satisfy
\begin{equation}
w_a \cdot w_b=c_{ab}
\end{equation}

Parallel to the pure gravity case, the single-centered attractor
flow is constructed as $S(\tau)=e^{K(\tau)}S_0$, where we choose
$K(\tau)$ to have the same properties as the generator $k$:
\begin{equation}
K^3(\tau)=0 \qquad \textrm{and} \qquad K^2(\tau) \,\,\textrm{rank
two}
\end{equation}
This determines $K(\tau)=k\tau+g$ where
\begin{equation}\label{kgB}
k=\sum_{a=1,2} [v_a w_a^T+ w_a v_a^T]S_0  \qquad \textrm{and}
\qquad g=\sum_{a=1,2}  [v_a m_a^T+ m_a v_a^T]S_0
\end{equation}where
the two 7-vectors $m_a$'s are orthogonal to $v_a$ and contain the
information of asymptotic moduli. Using $[[k,g],g]=0$, the current
is reduced to
\begin{equation}
J=\frac{S_0(k+\frac{1}{2}[k,g])S_0}{r^2}\hat{\vec{r}}
\end{equation}
from which we obtain $v_a$ and $w_a$ in terms of the charges and
the asymptotic moduli.

\section{Flow Generators in the $G_{2(2)}/(SL(2,\mathbb{R})\times SL(2,\mathbb{R}))$ Model}

We now explicitly construct the generators of single-centered
attractor flows. We start with the BPS flow which is associated
with a specific combination of the coset algebra generators
$a_{\alpha A}$. It can be derived from the condition of
preservation of supersymmetry. We then construct the non-BPS
attractor flow generator in analogy with the BPS one. In Section
5.2, we write $k_{BPS}$ and $k_{nonBPS}$ in terms of the $v_a$ and
$w_a$ vectors.  This form will be especially helpful in
generalizing to the multi-centered case.

\subsection{Construction of Flow Generators}

\subsubsection{Constructing $k_{BPS}$ using supersymmetry}

To describe BPS trajectories it is useful to remember that
the stabilizer of $S$ in $G_{2(2)}$ is $SO(1,2) \times SO(1,2)$,
corresponding to the elements of $G_{2(2)}$ which are
antisymmetric after multiplication by $S_0$. Geodesics are
exponentials of elements that are symmetric after multiplication
by $S_0$. Such elements sit in a $\mathbf{(2,4)}$ representation
of $SO(1,2) \times SO(1,2)$. A BPS trajectory is highest weight
for the first $SO(1,2)$. Labelling the symmetric generators as
$a_{\alpha A}$ under the two $SO(1,2)$ groups, a BPS trajectory is
generated by
\begin{equation}
k_{BPS}=a_{\alpha A} C^A z^{\alpha} \ .
\end{equation}
The twistor $z$ and the coefficients $C^A$ are fixed in terms of the
charges of the extremal BPS black hole and the condition of zero
time-like NUT charge.

To see why this is true, expand the coset element $k_{BPS}$ that
generates the BPS attractor flow using $a_{\alpha A}$:
\begin{equation}
k_{BPS}=a_{\alpha A}C^{\alpha A}
\end{equation}
where $C^{\alpha A}$ are conserved along the flow. On the other
hand, the conserved currents in the homogeneous space are
constructed by projecting the one-form valued Lie algebra
$g^{-1}\cdot dg$ onto $\mathbf{k}$, which gives the vielbein in
the symmetric space:
\begin{equation}
g^{-1} dg|_{\mathbf{k}} =  a_{\alpha A}V^{\alpha A}
\end{equation}
where $V^{\alpha A}$ is conserved:
\begin{equation}
\frac{d}{d\tau}\left(V_a^{\alpha A} \dot{\phi}^a \right) = 0
\end{equation}
Therefore, the expansion coefficients of $k_{BPS}$ are
\begin{equation}
C^{\alpha A} = V_a^{\alpha A} \dot{\phi}^a
\end{equation}
In terms of the vielbein, the supersymmetry condition that gives
the BPS geodesics are written as \cite{Pioline:2006ni}:
\begin{equation}
V^{\alpha A}z_{\alpha}=0
\end{equation}
That is:
\begin{equation}
V_a^{\alpha A} \dot{\phi}^a z_{\alpha} =0 \qquad \Longrightarrow
\qquad C^{\alpha A} z_{\alpha} = 0
\end{equation}
Define $z^{\alpha}=\epsilon^{\alpha\beta}z_{\beta}$,
\begin{equation}
C^{\alpha A}=C^Az^{\alpha}
\end{equation}
Therefore, the coset element $k_{BPS}$ is expanded by the coset
algebra basis $a_{\alpha A}$ as $k_{BPS}=a_{\alpha
A}C^Az^{\alpha}$.

Note that $k_{BPS}$ has five parameters $(C^{A},z)$ where $A=1,\dots,4$. As
will be shown later, $z$ can actually be determined by $(C^{A})$
and moduli at infinity. So the geodesic generated by $k_{BPS}$ is
indeed a four-parameter family. It is easy to show that $k_{BPS}$
is null, but more importantly, it is nilpotent:
\begin{equation}
k_{BPS}^3=0
\end{equation}
As will be shown later, $k_{BPS}$ indeed gives the correct BPS
attractor flow.

\subsubsection{Constructing $k_{NonBPS}$}

To construct the non-BPS attractor flow, one needs to find an
element in the coset algebra distinct from $k_{BPS}$ that
satisfies:
\begin{equation}
k^3_{NonBPS}=0
\end{equation}
The hint again comes from the BPS generator. Note that
$k_{BPS}=a_{\alpha A}P^{A}z^{\alpha}$ can be written as:
\begin{equation}\label{twistingBPS}
k_{BPS}=e^{-zL^{-}_{h}}k^0_{BPS}e^{zL^{-}_{h}}
\end{equation}
where $k^0_{BPS}$ spans only the right four coset generators
$a^{1A}$:
\begin{equation}
k^0_{BPS}=a_{1A}C^A
\end{equation}
That is, $k_{BPS}$ is generated by starting with the element
spanning the four generators annihilated by the horizontal $SL(2)$
raising operator $L^{+}_{h}$, then conjugating with the horizontal
$SL(2)$ lowering operator $L^{-}_{h}$. And it is very easy to show
that $(k^0_{BPS})^3=0$ which proves $(k_{BPS})^3=0$.

In $G_{2(2)}/SL(2,\mathbb{R})^2$, there are two third-degree
nilpotent generators in total. And since there are only two
$SL(2,\mathbb{R})$'s inside $\mathbf{H}$, a natural guess for a
non-BPS solution is to look at vectors with fixed properties under
the second $SL(2,\mathbb{R})$ group. An interesting condition is
to have positive charge under some rotation of $L^{3}_{v}$, i.e.
an $SL(2,\mathbb{R})$ rotation of $\sum_{A=1,2} a_{ \alpha A} C^{
\alpha A}$. Therefore, this suggests to us to start with the element
spanning the four generators annihilated by the square of the vertical
$SL(2,\mathbb{R})$ raising operator $(L^{+}_{v})^2$ and then conjugate
with the vertical $SL(2,\mathbb{R})$ lowering operator
$L^{-}_{v}$:
\begin{equation}
k_{NonBPS}(z)=e^{-zL_{v}^{-}}k^0_{NonBPS}e^{zL_{v}^{-}}
\end{equation}
where
\begin{equation}
k^0_{NonBPS}=a_{\alpha a}C^{\alpha a} \qquad \textrm{where}\qquad
\alpha, a=1,2.
\end{equation}
And one can show that: $(k^0_{NonBPS})^3=0$ which proves
$(k_{NonBPS})^3=0$. Moreover, $(k_{NonBPS})^2$ is rank two.

As long as one can pick the coefficients $C^{\alpha A}$ and the
twistor $z$ that describes the $SO(1,2)$ direction to be such that
the time NUT charge is zero, this generator will give nice non-BPS
extremal black holes. All the known non-BPS solutions may be
recovered this way, and more, as this construction gives absolute
freedom to pick the charges and moduli at infinity for the black
hole (clearly for certain values of charges and moduli the
solution will crash into a naked singularity, but this is to be
expected from comparison with the BPS case)

\subsection{Properties of Flow Generators}

\subsubsection{Properties of $k_{BPS}$}

We now turn to solving for $v_a$ and $w_a$ in (\ref{kInTermsOfvw}) in
terms of $C^A$ and $z$. First, from (\ref{kSquaredNullVectors}) we
know that the null space of $k^2$ is five-dimensional and the
$v_a$ span the two-dimensional complement of this null space. For
$k_{BPS}=a_{\alpha A}C^Az^{\alpha}$ the null space of
$(k_{BPS})^2$ does not depend on $C^A$. Therefore, the $v_a$
depend only on the twistor $z=z^2/z^1$.

Recall that we are using the basis where $k$ has the form
(\ref{kBasis}). From inspection of $k_{BPS}^2$, we find that
$(v_1, v_2)$ can always be chosen to have the form:\footnote{When
solving for $(v,w)$, there are some freedom on the choice of
$(v_1,v_2)$ and $(w_1,w_2)$: firstly, a rotational freedom
\begin{equation}
(v_1,v_2)\rightarrow(v_1,v_2)\left(\begin{array}{cc}
  R_{11} & R_{12}\\
  R_{21} & R_{22}
\end{array}\right) \qquad \textrm{and}\qquad (w_1,w_2)\rightarrow(w_1,w_2)\left(\begin{array}{cc}
  R_{11} & R_{12}\\
  R_{21} & R_{22}
\end{array}\right)
\end{equation}
where $R$ is orthogonal: $RR^{T}=1$ Secondly, a rescaling freedom:
\begin{equation}
v_a \rightarrow r v_a \qquad \textrm{and} \qquad w_a \rightarrow
\frac{1}{r} w_a
\end{equation}
}
\begin{equation}
v_1=(V_1,-\eta_1V_1,0) \qquad v_2=(-V_2,\eta_1 V_2,\sqrt{2})
\end{equation}
where $\eta_1$ is a 3d metric of signature $(1,-1,-1)$, and
$V_1,V_2$ are two three-vectors with
\begin{equation}\label{V12condition}
V_1\cdot V_1=0 \qquad V_1\cdot V_2=0 \qquad V_2\cdot V_2=-1
\end{equation}
Since any linear combination of $(v_1,v_2)$ forms a new set of
$(v_1,v_2)$, this means in particular that any $v_2+ c v_1$ gives
a new $v_2$. Looking at the forms of $(v_1, v_2)$, we see that
$V_2$ is defined up to a shifting of $V_1$ as $V_2=V^0_{2}-c V_1$.

An explicit computation shows that $V_1$ and $V_2$ are given by
the twistor $z$ and $u$ as
\begin{equation}
\label{AandBVectorForm}
V_1 =\left(\begin{array}{c} (z^1)^2+(z^2)^2\\
(z^1)^2-(z^2)^2\\
2z^1z^2
\end{array} \right)\qquad \qquad V_2
=\frac{1}{z^1u^2-z^2u^1}\left(\begin{array}{c} z^1u^1+z^2u^2\\
z^1u^1-z^2u^2\\
z^1u^2+z^2u^1
\end{array} \right).\end{equation}
where the twistor $u=\frac{u^2}{u^1}$ is related to $c$ by
\begin{equation}
u=-\frac{1+2cz}{1-2cz}z
\end{equation}
The twistor representation \footnote{With the inner product of
three-vectors defined as
\begin{equation}
v_a\cdot v_b= v^T_a \eta_1 v_b
\end{equation}
The twistor representation of a three-vector $v=(x,y,z)$ is
\begin{equation}
\sigma_{v}=x \sigma_0+y\sigma_3+z\sigma_1=\left(\begin{array}{cc} x+y&z\\
z&x-y\end{array} \right)
\end{equation}
It's length is
\begin{equation}
v^T\eta_1 v=\det(\sigma_v)=x^2-y^2-z^2
\end{equation}} of $V_1$ and $V_2$ are
\begin{equation}
\label{ABSpinorForm} V_1^{\alpha\beta}=2z^{\alpha}z^{\beta} \qquad
\qquad V_2^{\alpha \beta} = z^{\alpha} u^{\beta} + z^{\beta}
u^{\alpha}
\end{equation}
where we have used the rescaling freedom to set $z^1u^2-z^2u^1$ to
be $1$.  Note that for the BPS case, the twistor $u$ is totally
arbitrary.

Now we solve for $w_a$. The condition that $w_a$ are orthogonal to
$v_a$ dictates that they have the form:
\begin{equation}\label{W12formBPS}
w_1 = (W_1,\eta_1 W_1,0) \qquad w_2 = (W_2, \eta_1 W_2,0)
\end{equation}
where $W_1$ and $W_2$ are linearly independent, and are related to
the charges by $w_a\cdot w_b=c_{ab}$:
\begin{equation}
W_1\cdot W_1=\frac{1}{2}c_{11} \qquad W_1\cdot
W_2=\frac{1}{2}c_{12} \qquad W_2\cdot W_2=\frac{1}{2}c_{22}
\end{equation}
Recall that $V_2$ is defined up to a shift by $V_1$:
$V_2=V^0_2-cV_1$. The consequence is that $W_1$ is defined up to a
shift by $W_2$: $W_1=W^0_1+c W_2$. Note that the numerical factors
in front of $V_1$ and $W_2$ are opposite. Write down $(W^0_1,W_2)$
in terms of $(C^A,z)$:
\begin{equation}
W^0_1 =\frac{1}{4z}\left(\begin{array}{c} (C^2+C^4)+(C^1+C^3)z\\
(C^2-C^4)+(C^1-C^3)z\\
2C^3+2C^2 z
\end{array} \right)\qquad W_2=\frac{1}{2}\left(\begin{array}{c} -(C^2+C^4)+(C^1+C^3)z\\
-(C^2-C^4)+(C^1-C^3)z\\
-2C^3+2C^2 z
\end{array} \right)
\end{equation}
The twistor representations of $W_1$ and $W_2$ are
\begin{equation}\label{CDspinorB}
W_1=\left(\begin{array}{cc} C^1u^2-C^2u^1&C^2u^2-C^3u^1\\
C^2u^2-C^3u^1&C^3u^2-C^4u^1
\end{array} \right)\qquad W_2=\left(\begin{array}{cc} C^1z^2-C^2z^1&C^2z^2-C^3z^1\\
C^2z^2-C^3z^1&C^3z^2-C^4z^1
\end{array} \right)
\end{equation}
Define the totally symmetric $P^{\alpha\beta\gamma}$:
\begin{equation}
P^{111}=C^1 \qquad P^{112}=C^2 \qquad P^{122}=C^3 \qquad
P^{222}=C^4
\end{equation}
Then the three-vectors $(W_1,W_2)$ span the four dimensional space
\begin{equation}
(W^{\alpha}_1,W^{\alpha}_2)_{BPS}=(P^{\alpha\beta\gamma}u_{\gamma},P^{\alpha\beta\gamma}z_{\gamma})
\end{equation}

\subsubsection{Properties of $k_{NonBPS}$}

The form of $v_a$ for the non-BPS case is only slightly different
from the BPS case: the two vectors $v_a$ can be chosen to have the
form:
\begin{equation}
v_1 = (V_1, \eta_1 V_1,0)  \qquad v_2 = (V_2,-\eta_1 V_2,\sqrt{2})
\end{equation}
where $V_1,V_2$ are two three-vectors satisfying the same
condition as the BPS ones (\ref{V12condition}). Again, the vectors
$V_1$ and $V_2$ can be written as (\ref{AandBVectorForm}), and the
twistor representations are given in (\ref{ABSpinorForm}) with one
major difference: $u$ is no longer arbitrary, but is determined by
$C^{\alpha A}$ as:
\begin{equation}
u=\frac{u^2}{u^1}=\frac{C^{22}}{C^{12}}
\end{equation}

The form of $(w_1,w_2)$ are also slightly different from the BPS
one (\ref{W12formBPS})
\begin{equation}\label{W12formNB}
w_1 = (W_1,- \eta_1 W_1,0) \qquad w_2 = (W_2, \eta_1 W_2,0)
\end{equation}
The $(W_1,W_2)$ can be written in terms of $(C^{\alpha a},z)$ thus:
\begin{equation}
W_1=\frac{1}{2(C^{22}z^1-C^{12}z^2)}\left(\begin{array}{c} [(C^{11}C^{22}-C^{12}C^{21})z^2+(C^{22})^2]+[C^{11}C^{22}-C^{12}C^{21}+(C^{12})^2]\\
 -[(C^{11}C^{22}-C^{12}C^{21})z^2+(C^{22})^2]+[C^{11}C^{22}-C^{12}C^{21}+(C^{12})^2]\\
2[(C^{11}C^{22}-C^{12}C^{21})z+C^{12}C^{22}]
\end{array} \right)
\end{equation}
\begin{equation}
W_2=-\frac{1}{2}\left(\begin{array}{c} z[C^{11}z^2+(3C^{12}-C^{21})z-2C^{22}]+[ C^{11}z+C^{12}-C^{21}]\\
 -z[C^{11}z^2+(3C^{12}-C^{21})z-2C^{22}]+[ C^{11}z+C^{12}-C^{21}]\\
2[C^{11}z^2+(2C^{12}-C^{21})z-C^{22}]
\end{array} \right)
\end{equation}
In terms of $u=\frac{u^2}{u^1}=\frac{C^{22}}{C^{12}}$, the twistor
representation of $W_1$ and $W_2$ are:
\begin{equation}
W_1^{\alpha\beta}=u^{\alpha}u^{\beta}+(C^{11}u^2-C^{12}u^1)z^{\alpha}z^{\beta}
\end{equation}
\begin{equation}
W_2^{\alpha\beta}=(z^{\alpha}u^{\beta}+u^{\alpha}z^{\beta})+(C^{21}-C^{11}z-3u^1)z^{\alpha}z^{\beta}
\end{equation}
As a consequence, the precise value of $u$ is an extra constraint
on the vectors $w_a$, and there is only a three-dimensional space
of them, with $(W_1,W_2)$ a linear combination of $(0,V_1)$,
$(V_1,0)$ and $(u^{\alpha} u^{\beta},2 V_2)$.
\begin{equation}
(W_1,W_2)_{NonBPS}=m (0,V_1)+  n (V_1,0)+ \ell (u^{\alpha}
u^{\beta},2 V_2)
\end{equation}

\section{Single-centered Attractor Flows in $G_{2(2)}/(SL(2,\mathbb{R})\times SL(2,\mathbb{R}))$ Model}

Now that we have completely characterized the generators of single-centered attractor flow, we can lift the geodesics to four-dimensional black hole solutions. After some calculational effort, we find that the BPS solution is given in terms of harmonic functions. Next, we show that the non-BPS case is qualitatively different, and the final solutions cannot be formulated so simply.

\subsection{BPS Attractor Flow}
\subsubsection{Lifting a geodesic to $4D$}

We have already noted that the flow starting from generic asymptotic moduli $(x_0, y_0)$
is generated by $M(\tau)=e^{(k\tau+g)/2}$, with $g$ defined before
in (\ref{kgB}). The matrix $g$ has the same form as $k$. Therefore, in the BPS case, it has the expansion
\begin{equation}
g_{BPS}=a_{\alpha A}z^{\alpha}G^{A}
\end{equation}
where the twistor $z$ is the same as the one in $k_{BPS}$. The
flow of $(x,y)$ and $u$ can be extracted from the symmetric matrix
$S(\tau)=M(\tau)S_0M(\tau)^T$ via (\ref{xyfromS}) and
(\ref{ufromS}). Using $k^3_{BPS}=0$ and $g^3=0$,
$S(\tau)=S_0e^{k\tau+g}$ is a quadratic function of $\tau$:
\begin{eqnarray}\label{Stau}
S_{55}(\tau)_{BPS}&=&\alpha_B(\tau) +\beta_B(\tau) -1 \non\\
S_{35}(\tau)_{BPS}&=&\gamma_B( \tau)+\delta_B (\tau) \\
S_{33}(\tau)_{BPS}&=&\epsilon_B (\tau)+\zeta_B (\tau)-1 \non
\end{eqnarray}
where $\{\alpha_{B}(\tau),\gamma_{B}(\tau),\epsilon_{B}(\tau)\}$
are quadratic functions of $\tau$, and
$\{\beta_{B},\delta_{B},\zeta_{B}\}$ are linear functions of
$\tau$:
\begin{eqnarray}
\alpha_B (\tau)&=& z^2(H^2 H^4-(H^3)^2)+z(H^2 H^3-H^1 H^4)+(H^1 H^3-(H^2)^2)\non\\
\beta_B(\tau) &=&(H^2-H^4)z+(H^3-H^1)\non\\
\gamma_B (\tau)&=& -\frac{1}{2}\left((H^1H^4-H^2 H^3) (z^2-1)+2(H^2(H^2+H^4)-H^3(H^1+H^3))z\right)\non\\
 \delta_B (\tau)&=& -\frac{1}{2}((H^1+H^3)z-(H^2+H^4))\non\\
\epsilon_B (\tau)&=&(H^1 H^3-(H^2)^2)z^2+(H^1 H^4-H^2 H^3)z+(H^2 H^4-(H^3)^2)\non\\
\zeta_B(\tau) &=& 2(H^2 z+H^3)
\end{eqnarray}
where $H^A$ is a linear function of $\tau$ defined as
$H^A(\tau)\equiv C^A \tau+G^A$.

The attractor values are reached when $\tau\rightarrow \infty$
along the geodesic:
\begin{equation}\label{moduli*kB}
x^{*}_{BPS}=-\frac{(k^2S_0)_{35}}{(k^2S_0)_{33}}\qquad
y^{*}_{BPS}=\frac{\sqrt{(k^2S_0)_{33}(k^2S_0)_{55}-(k^2S_0)^2_{35}}}{(k^2S_0)_{33}}
\end{equation}
and \begin{equation}\label{umoduli*kB}
u^{*}_{BPS}=\frac{1}{\sqrt{(k^2S_0)_{33}(k^2S_0)_{55}-(k^2S_0)^2_{35}}}
\end{equation}
The asymptotic moduli $(x_0,
y_0)$ can be expressed in terms of $(G^A,z)$ by extraction from
$S=e^{g}S_0$.

\begin{figure}[htbp]
  \centering
  \includegraphics{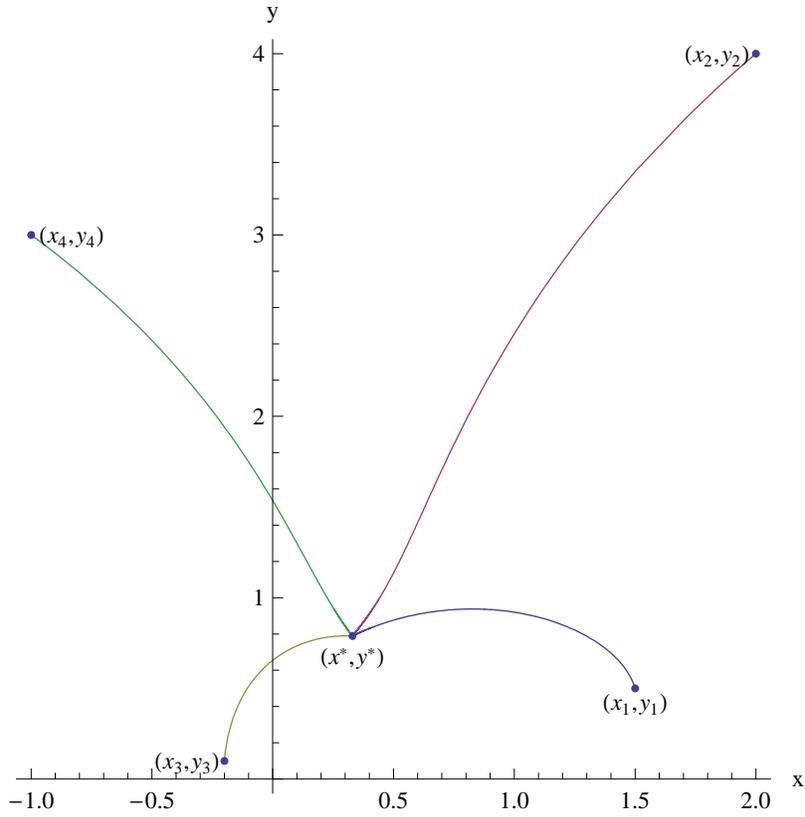}
  \caption{Sample BPS flow. The attractor point is labeled $(x^*,y^*)$. The initial points of each flow are given by  $(x_1=1.5, y_1=0.5), (x_2=2, y_2=4), (x_3=-0.2, y_3=0.1), (x_4=-1, y_4=3)$}
  \label{fig:BPSflow}
\end{figure}

Using this technique, one can construct all BPS single-centered black holes. The charges of each black hole can be read off from the current $J$ using (\ref{ExtractCharges}). One example is given in Figure \ref{fig:BPSflow}, where we parametrically plot $x(\tau)$ and $y(\tau)$ for a BPS black hole with charges $(p^0, p^1, q_1, q_0)= (5, 2,7,-3)$ and attractor point $(x^*,y^*)=(0.329787,0.788503)$.
\footnote{The discriminant of the charge $(5,2,7,-3)$ is positive, so this is indeed a BPS solution.}

\subsubsection{$4D$ solution for given set of charges}

To get the solution for a specific set of charges requires more effort. In this section, we present the analytical result for any set of charges $(p^I, q_I)$.

The ten parameters in $k_{BPS}$ and $g$ are $\{z, u, C^A, G^A\}$,
among which the twistor $u$ is arbitrary, corresponding to the
freedom of the shift by $(v_1, w_2)$ in the definition of
$(v_2,w_1)$: $(v_2, w_1)\rightarrow (v_2+c v_1, w_1-c w_2)$. The
remaining true parameters are enough to parameterize the four
D-brane charges $(p^I,q_I)$ and the arbitrary asymptotic moduli
$(x_0, y_0)$ under the condition of vanishing Taub-NUT charge and
fixing $u_0=1$. We now solve for $k_{BPS}$ and $g$ for the given
D-brane charges and  $(x_0, y_0)$, using the eight constraints,
namely, $4$ charges and zero Taub-NUT charge plus $3$ asymptotic
moduli, to fix $C^A$ and $G^A$, leaving the other twistor $z$
unfixed.\footnote{The twistor $z$ can be left unfixed because we
will not specify the asymptotic values of the scalars with
translational invariance, namely, $(\{a,A^I,B_I\})$. Fixing them
can fix the twistor $z$.}

For the sake of simplicity, we will denote $k_{BPS}$ by $k$ for the rest of this section. Then the current $J(Q)=\frac{k^T}{r^2}$
gives the five coupled equations:
\begin{equation}\label{kfromQsingleg}
\mathbf{Q}=S_0 (k+\frac{1}{2}[k, g])S_0
\end{equation}
In order to show that the BPS flow can be expressed in terms of
harmonic functions:
\begin{equation}
H(\tau)=Q\tau+h \qquad \textrm{with} \qquad Q=(p^I, q_I) \qquad
\textrm{and} \qquad  h=(h^I, h_I)
\end{equation}
we will solve $g$ in terms of $h$ instead of $(x_0, y_0)$. The
four $h$'s relate to the asymptotic moduli by
\begin{equation}\label{x0y0h}
x_0=x^{*}(Q \rightarrow h)  \qquad y_0=y^{*}(Q \rightarrow h)
\qquad u_0=u^{*}(Q \rightarrow h)
\end{equation}
and there is one extra degree of freedom to be fixed later.

To evaluate $[k,g]$, we first use the commutation relation
$(\ref{aacommu})$ to obtain
\begin{equation}\label{CGcommu1}
[a_{1A}C^A,a_{1B}G^{B}]=\langle C,G \rangle(-4L^{+}_{h})
\end{equation} where the product between
$C^A$ and $G^A$ is defined as $\langle C,G \rangle \equiv C^1
G^4-3C^2 G^3+3C^3 G^2-C^4 G^1$. Then twisting Eq (\ref{CGcommu1})
with the twistor $z$ as in (\ref{twistingBPS}) gives the
commutator of $k$ and $g$ with the same twistor $z$:
\begin{equation}
[k,g]=[a_{\alpha A}z^{\alpha} C^A,a_{\beta B}z^{\beta}
G^B]=\langle C,G \rangle \Theta
\end{equation}
where $\Theta$ is defined as $\Theta \equiv
-\frac{4}{1+z^2}e^{-zL^{-}_{h}}L^{+}_he^{zL^{-}_{h}}$. On the
other hand, using (\ref{kgB}),
\begin{equation}
[k,g]=(v_2v^T_1-v_1v^T_2)S_0(w_2\cdot m_1-w_1\cdot m_2)
\end{equation}
$\Theta$ can also be written as $\Theta=(v_2v^T_1-v_1v^T_2)S_0$,
and we can check that $(w_2\cdot m_1-w_1\cdot m_2)=\langle C,G
\rangle$.

First, separate from $G^A$ the piece which has the same dependence
on $(h,z)$ as $C^A$ on $(Q,z)$:
\begin{equation}
G^{A}=G^A_h+E^A \qquad \textrm{with} \qquad G^{A}_h\equiv C^A(Q
\rightarrow h, z)
\end{equation}
That is, $g$ contains two pieces:
\begin{equation}\label{gghlambda}
g=g_h+\Lambda \qquad \textrm{with} \qquad g_{h}=a_{\alpha
A}z^{\alpha}G^A_{h} \qquad \textrm{and} \qquad \Lambda=a_{\alpha
A}z^{\alpha}E^A
\end{equation}
We need to solve for $E^A$.

There are three constraints from (\ref{x0y0h}). The $(x_0, y_0)$
and $u_0$ are extracted from the symmetric matrix $S=e^{g}S_0$ via
(\ref{xyfromS}) and $(\ref{ufromS})$. On the other hand, requiring
(\ref{x0y0h}) gives $(x_0, y_0, u_0)$ in terms of $h$:
\begin{equation}\label{moduli*gB}
x_0=-\frac{(g_h^2S_0)_{35}}{(g_h^2S_0)_{33}}\qquad
y_0=\frac{\sqrt{(g_h^2S_0)_{33}(g_h^2S_0)_{55}-(g_h^2S_0)^2_{35}}}{(g_h^2S_0)_{33}}
\qquad
u_0=\frac{1}{\sqrt{(g_h^2S_0)_{33}(g_h^2S_0)_{55}-(g_h^2S_0)^2_{35}}}
\end{equation}
Therefore, defining $\Pi\equiv(e^{g}-\frac{g^2_h}{2})S_0$, $\Pi$
has to satisfy three constaints:
\begin{equation}\label{Econstraint123}
\Pi_{33}=\Pi_{35}=\Pi_{55}=0
\end{equation}
in order for (\ref{moduli*gB}) to hold for arbitrary $h$. Using the unfixed
degree of freedom in $h$'s to set $\langle C,G_h \rangle=0$,
(\ref{kfromQsingleg}) becomes
\begin{equation}\label{kfromQsinglelambda}
\mathbf{Q}=S_0 (k +\frac{1}{2}[k, \Lambda])S_0
\end{equation}
The zero Taub-NUT charge condition in (\ref{kfromQsinglelambda})
imposes the fourth constraint on $\Lambda$: the (3,2)-element of
$S_0 (k +\frac{1}{2}[k, \Lambda])S_0$ for arbitrary $k$ has to
vanish. Combining with (\ref{Econstraint123}), we have $4$
constraints to fix $E^A$ to be:
\begin{equation}\label{Esolution}
E^1=-E^3=-\frac{1}{1+z^2} \qquad E^2=-E^4=\frac{z}{1+z^2}
\end{equation}

The remaining $4$ conditions in the coupled equations
(\ref{kfromQsinglelambda}) determine $C^A$ in the BPS generator
$k_{BPS}=a_{\alpha A}z^{\alpha}C^{A}$ to be
\begin{eqnarray}\label{CfromQzBgenericz}
C^1&=&\sqrt{2}\frac{-q_0-q_1z-3p^1z^2+p^0z^3}{(1+z^2)^2}\non\\
C^2&=&\sqrt{2}\frac{-\frac{q_1}{3}-(2p^1-v_0)z+(p^0+2\frac{q_1}{3})z^2+p^1z^3}{(1+z^2)}\non\\
C^3&=&\sqrt{2}\frac{-p^1+(p^0+2\frac{q_1}{3})z+(2p^1-v_0)z^2-\frac{q_1}{3}z^3}{(1+z^2)}\non\\
C^4&=&\sqrt{2}\frac{p^0+3p^1z-q_1z^2+q_0z^3}{(1+z^2)^2}
\end{eqnarray}
The $G_h^A$ are then determined by $G^A_h=a_{\alpha
A}z^{A}C^A(Q\rightarrow h,z)$. Using the solution of $C^A$ and
$G_h^A$, we see the product $\langle C^A, G^A_h \rangle$ is
proportional to the symplectic product of $(p^I,q_I)$ and $(h^I,
h_I)$:
\begin{equation}
\langle C^A, G^A_h \rangle=\frac{2}{1+z^2}<Q,h>\qquad
\textrm{where} \qquad <Q,h>\equiv p^0h_0+p^1h_1-q_1h^1-q_0h^0
\end{equation}
The condition $\langle C^A, G^A_h \rangle=0$ is then the
integrability condition on $h$:
\begin{equation}\label{Integrability}
<Q,h>=p^0h_0+p^1h_1-q_1h^1-q_0h^0=0
\end{equation}

Substituting the expressions of $C^{A}$ and $G^A$ in terms of $(p^I,
q_I)$ into (\ref{Stau}), we obtain the BPS attractor flow in terms
of the charges $(p^I, q_I)$. In particular, the attractor values are
\begin{equation}\label{moduli*QB}
x^{*}_{BPS} = -\frac{p^0 q_0+p^1 \frac{q_1}{3}}{2[(p^1)^2+p^0
\frac{q_1}{3}]} \qquad y^*_{BPS} =
\frac{\sqrt{J_4(p^0,p^1,\frac{q_1}{3}, q_0)}}{2[(p^1)^2+p^0
\frac{q_1}{3}]}
\end{equation}
where $J_4(p^0,p^1,q_1, q_0)$ is the quartic $E_{7(7)}$ invariant:
\begin{equation}
J_4(p^0,p^1,q_1, q_0)=3 (p^1q_1)^2 -6(p^0q_0) (p^1
q_1)-(p^0q_0)^2-4(p^1)^3 q_0+ 4 p^0(q_1)^3
\end{equation}
thus $J_4(p^0,p^1,\frac{q_1}{3}, q_0)$ is the discriminant of charge.
The attractor values match those from the compactification of Type II string theory
on diagonal $T^6$, with $q_1 \rightarrow \frac{q_1}{3}$. The
attractor value of $u$ is
\begin{equation}\label{umoduli*QB}
u^*_{BPS} = \frac{1}{\sqrt{J_4(p^0,p^1,\frac{q_1}{3}, q_0)}}
\end{equation}
The constraint on $h$ from $u_0=1$ is then
$J_4(h^0,h^1,\frac{h_1}{3},h_0)=1$.

Now we will show that the geodesic we constructed above indeed
reproduces the attractor flow given by replacing charges by the
corresponding harmonic functions in the attractor moduli. Using
the properties of $\Lambda$, we have proved that, in terms of $k$
and $g$, the flow of $(x,y)$ can be generated from the attractor
value by replacing $k$ with the harmonic function $k\tau+g_h$:
\begin{equation}\label{kharmonic}
x(\tau)=x^{*}(k\rightarrow k\tau+g_h)\qquad
y(\tau)=y^{*}(k\rightarrow k\tau+g_h)
\end{equation}
Since $k$ and $g_h$ have the same twistor $z$, this is equivalent
to replacing the $C^A$ with the harmonic function $C^A\tau+G^A_h$
while leaving the twistor $z$ fixed:
\begin{equation}\label{Charmonic}
x(\tau)= x^{*}(C^A\rightarrow C^A\tau+G^A_h,z)\qquad
y(\tau)=y^{*}(C^A\rightarrow C^A\tau+G^A_h,z)
\end{equation}
Since $C^A$ is linear in $Q$ and $G^A_h$ linear in $h$, and
since $z$ drops out after plugging in the solution of $C^A$ in
terms of $(Q,z)$ and $G^A_h$ in terms of $(h,z)$, this proves that
the flow of $(x_0, y_0)$ is given by replacing the charges in the
attractor moduli by the corresponding harmonic functions:
\begin{equation}\label{QharmonincProcedure}
x_{BPS}(\tau)=x^{*}_{BPS}(Q\rightarrow Q\tau+h)\qquad
y_{BPS}(\tau)=y^{*}_{BPS}(Q\rightarrow Q\tau+h)
\end{equation}
The integrability condition $<Q, h>=0$, in terms of $H=Q\tau+h$,
is
\begin{equation}
< H,dH>=0
\end{equation}

\subsection{non-BPS Attractor Flow }

\subsubsection{Lifting a geodesic to $4D$}

Similar to the BPS attractor flow, the non-BPS flow is generated by
$M(\tau)=e^{(k\tau+g)/2}$, and $(x,y)$ can be extracted from the
symmetric matrix $S(\tau)$ by (\ref{xyfromS}) and the relevant
elements of $S(\tau)$ are given by (\ref{Stau}). The only
difference is that now
$\{\alpha_{B},\beta_{B},\gamma_{B},\delta_{B},\epsilon_{B},\zeta_{B}
\}$ are changed into the non-BPS counterparts
$\{\alpha_{NB},\beta_{NB},\gamma_{NB},\delta_{NB},\epsilon_{NB},\zeta_{NB}
\}$, which can be written in terms of $H^{\alpha a}\equiv
C^{\alpha a}\tau+G^{\alpha a}$ and $z$:
\begin{eqnarray}\label{abcdefNB}
\alpha_{NB}(\tau) &=& -((H^{21}H^{12}-H^{22}H^{11})(z^2-1)^2+(H^{22})^2z^2-2zH^{12}H^{22}+(H^{12})^2)\non\\
\beta_{NB} (\tau)&=& (z^2-1)(H^{11}-H^{21}z)-3z^2H^{22}+2zH^{12}+H^{22}\non\\
\gamma_{NB}(\tau) &=& (z^2-1)(2z(H^{11}H^{22}-H^{12}H^{21})+H^{22}H^{12})+z((H^{22})^2-(H^{12})^2)\non\\
\delta_{NB}(\tau) &=& \frac{1}{2}(z(1+z^2)H^{11}+z^2(3H^{12}-H^{21})-2H^{22}z+(H^{12}-H^{21})) \non\\
\epsilon_{NB} (\tau)&=& (4H^{11}H^{22}-H^{12}(H^{12}+4H^{21}))z^2-2H^{12}H^{22}z-(H^{22})^2\non\\
\zeta_{NB}(\tau) &=& 2(H^{11}z^2+(2H^{12}+H^{21})z+H^{22})
\end{eqnarray}
Note that $\frac{H^{22}}{H^{12}}=u$ is fixed, independent of
$\tau$. The non-BPS flow written in terms of $(H^{\alpha a},z)$
has the same simple form as the BPS flow, i.e. the scalars are rational
functions with both the numerator and denominator being only
quadratic. This is due to the nilpotency of the generator: $k^3=0$. Again,
the attractor values are reached when $\tau \rightarrow \infty$,
and the asymptotic moduli can be expressed in terms of $(G^{\alpha
a},z)$ by extraction from $S=e^{g}S_0$.

Unlike the BPS case, there are only eight parameters in
$k_{NonBPS}$ and $g_{NonBPS}$: the two twistors $(z,u)$ and
$(C^{\alpha a},G^{\alpha a})$ under the constraints that
\begin{equation}
u=\frac{C^{22}}{C^{12}}=\frac{G^{22}}{G^{12}}
\end{equation}
Therefore, while $k_{BPS}$ and $g_{BPS}$ can parameterize
arbitrary $(p^I, q_I)$ and $(x_0, y_0)$ while leaving $(z,u)$
free, all the parameters in $k_{NonBPS}$ and $g_{NonBPS}$,
including $(z,u)$, will be fixed.

\begin{figure}[htbp]
  \centering
  \includegraphics{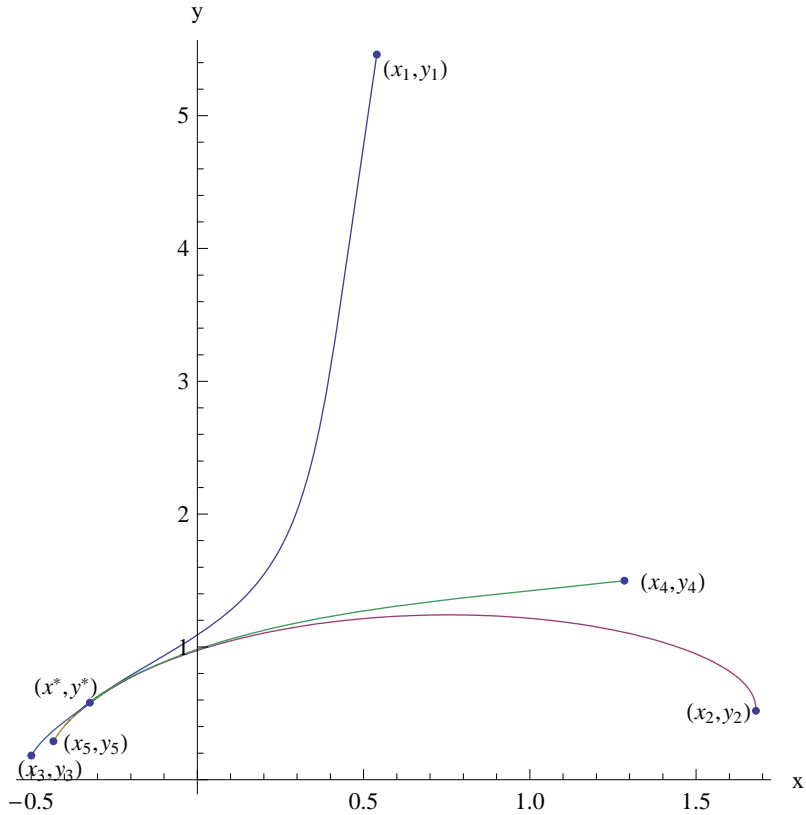}
  \caption{Sample non-BPS flow. The attractor point is labeled $(x^*,y^*)$. The initial points of each flow are given by: $(x_1=0.539624, y_1= 5.461135), (x_2=1.67984, y_2=0.518725), (x_3=-0.432811,y_3=0.289493),(x_4=1.28447, y_4=1.49815), (x_5=-0.499491, y_5=0.181744)$}
  \label{fig:NonBPSflow}
\end{figure}

The attractor point in terms of $C^{\alpha a}$ is
\begin{equation}\label{moduli*CNBsingle}
x^{*}_{NonBPS}=\frac{\gamma_{NB}(H^{\alpha a}\rightarrow C^{\alpha
a})}{\epsilon_{NB}(H^{\alpha a}\rightarrow C^{\alpha a})} \qquad
y^{*}_{NonBPS}=\frac{\sqrt{\alpha_{NB}(H^{\alpha a}\rightarrow
C^{\alpha a})\epsilon_{NB}(H^{\alpha a}\rightarrow C^{\alpha
a})-\gamma^2_{NB}(H^{\alpha a}\rightarrow C^{\alpha
a})}}{\epsilon_{NB}(H^{\alpha a}\rightarrow C^{\alpha a})}
\end{equation}
with $z$ given by
\begin{equation}
 \frac{1}{2} \left(-3 C^{12}-C^{21}+z\left((z^2-3) C^{11}+3z(C^{12}+C^{21})+6C^{22}\right)\right) = 0
\end{equation}
As in the BPS case, the charges of the black hole are read off from the current $J$ using (\ref{ExtractCharges}).  We have checked that the attractor point is a
non-supersymmetric critical point of the black hole potential
$V_{BH}=|Z|^2+|DZ|^2$:
\begin{equation}
\partial V_{BH}=0 \qquad \textrm{and} \qquad DZ\neq 0
\end{equation}
It reproduces the results reported in the literature
\cite{Tripathy:2005qp}. An example of the non-BPS attractor flow is shown in Figure \ref{fig:NonBPSflow}, with  $(p^0, p^1, q_1, q_0)= (5, 2, 7, 3)$ and attractor point $(x^*,y^*)=(-0.323385, 0.580375)$. Note that $J_4(5, 2, 7/3, 3)<0$, so this is indeed a non-BPS black hole. Unlike the BPS attractor flow, all the non-BPS flows starting from different asymptotic moduli have the same tangent direction at the attractor point. The mass matrix of the black-hole potential at a BPS critical point has two identical eigenvalues, whereas the eigenvalues at a non-BPS critical point are different. The common tangent direction for the non-BPS flows corresponds to the eigenvector associated with the smaller mass.

\subsubsection{$4D$ solution for given set of charges}

We now discuss how to construct the non-BPS black hole solution for a specific set of charges $(p^I, q_I)$.

One major difference between the non-BPS case and the BPS case is that
\begin{equation}[k_{NonBPS},g_{NonBPS}]=0\end{equation}
automatically, since the
forms of $(w_1,w_2)$ and $(m_1, m_2)$ guarantee that $w_1\cdot
m_2=w_2 \cdot m_1=0$. Thus the charge equation
(\ref{kfromQsingleg}) becomes simply
\begin{equation}\label{kfromQsingleNB}
\mathbf{Q}_{NonBPS}=S_0 (k_{NonBPS})S_0
\end{equation}
These five coupled equations determine the two twistors $(z,u)$
and $C^{\alpha a}$ in terms of $(p^I, q_I)$. Similar to the BPS case, the four equations which determine the D-brane charge allow us to write $C^{\alpha a}$ in terms of the charges $(p^0,p^1,q_1,
q_0)$ and the twistor $z$ via
\begin{eqnarray}\label{CfromQzNB}
C^{11}&=&\frac{(-2q_0+6(p^1-q_0)z^2+4(p^0+q_1)z^3-6p^1z^4)}{\sqrt{2}(1+z^2)^3}\non\\
C^{12}&=&\frac{(p^0+\frac{q_1}{3})-2(2p^1-q_0)z-(p^0+5\frac{q_1}{3})z^2+2p^1z^3}{\sqrt{2}(1+z^2)^2}\non\\
C^{21}&=&\frac{(p^0-q_1)-4q_1z^2+4(3p^1-q_0)z^3+(3p^0+q_1)z^4}{\sqrt{2}(1+z^2)^3}\non\\
C^{22}&=&\frac{2p^1+(p^0+5\frac{q_1}{3})z-2(2p^1-q_0)z^2-(p^0+\frac{q_1}{3})z^3}{\sqrt{2}(1+z^2)^2}
\end{eqnarray}
and $u=\frac{C^{22}}{C^{12}}$. In contrast to the BPS case, the $G^{\alpha a}$ do not enter the equations and therefore cannot be used to eliminate the twistor $z$. Requiring the Taub-NUT charge to vanish gives the following degree-six equation for the $z$:
\begin{equation}\label{zfromQNB}
p^0z^6+6p^1z^5-(3p^0+4q_1)z^4-4(3p^1-2q_0)z^3+(3p^0+4q_1)z^2+6p^1z-p^0=0
\end{equation}
 The three parameters in
$g_{NonBPS}$, namely, $G^{\alpha a}$ with the constraint
$\frac{G^{22}}{G^{12}}=u$ are then fixed by the given asymptotic
moduli $(x_0, y_0)$ and $u_0=1$.

Similar to the BPS flow, the full non-BPS flow can be generated
from the attractor value by replacing $C^{\alpha a}$ with the
harmonic function $H^{\alpha a}(\tau)=C^{\alpha a}\tau+G^{\alpha
a}$, while leaving $z$ unchanged as in (\ref{Charmonic}). However,
there are two major differences. First, the harmonic functions
$H^{\alpha a}$ have to satisfy the constraint
\begin{equation}\label{NBHFconstraint}
\frac{H^{22}(\tau)}{H^{12}(\tau)}=u=\frac{C^{22}}{C^{12}}=\frac{G^{22}}{G^{12}}
\end{equation} Note that this does not impose
any constraint on the allowed asymptotic moduli since there are
still three degrees of freedom in $G^{\alpha a}$ to account for
$(x_0, y_0, u_0)$. We will see later that it instead imposes a stringent
constraint on the allowed D-brane charges in the multi-centered
non-BPS solution.

Secondly, unlike the BPS flow, replacing $C^{\alpha a}$ in the
attractor moduli by the harmonic function $H^{\alpha a}(\tau)$ is
not equivalent to replacing the charges $Q$ with $H=Q\tau+h$ as in
 (\ref{QharmonincProcedure}). The twistor $z$ here is
no longer free, but is determined in terms of the charges as a
root of the degree-six equation (\ref{zfromQNB}), so replacing $Q$
by $Q\tau+h$, for generic $Q$ and $h$, would not leave $z$
invariant. Therefore, the generic non-BPS flow cannot be given by
the naive harmonic function procedure, as proposed by Kallosh et
al \cite{Kallosh:2006ib}. Next, we will define the subset of the
NonBPS single-centered flow that can be constructed by the
harmonic function procedure.

When the attractor has only $D4-D0$ charges, namely,
$Q_{40}=(0,p^1, 0,q_0)$, (\ref{zfromQNB}) has a root $z=0$, which
is independent of the value of charges. If the asymptotic moduli
$h$ is also of the form of $h_{40}=(0,h^1, 0,h_0)$, replacing
$Q_{40}$ by $Q_{40}\tau+h_{40}$ would leave the solution $z=0$
invariant. Now we will use the duality symmetry to extend the
subset to a generic charge system with restricted asymptotic
moduli.

The one-modulus system can be considered as the STU attractor with
the three moduli $(S,T,U)$ identified. Since the STU model has
$SL(2,\mathbb{Z})^3$ duality symmetry at the level of the
equations of motion, the one-modulus system has an
$SL(2,\mathbb{Z})$ duality symmetry coming from identifying the
three $SL(2,\mathbb{Z})$ symmetries of the STU model. That is, the one-modulus system is
invariant under the following element of $SL(2,\mathbb{Z})^3$
\begin{equation}\label{STUtransform}
\hat{\Gamma}=\left(\begin{array}{cc} a & b  \\
c & d  \end{array} \right)\otimes \left(\begin{array}{cc} a & b  \\
c & d  \end{array} \right)\otimes \left(\begin{array}{cc} a & b  \\
c & d  \end{array} \right) \qquad \textrm{with} \qquad ad-bc=1
\end{equation}
The modulus $z=x+i y$ transforms as
\begin{equation}
z\rightarrow \hat{\Gamma}z=\frac{a z+b}{c z+d}
\end{equation}
and the transformation on the charges is given by
\cite{Behrndt:1996hu}. A generic charge $(p^0,p^1,q_1, q_0)$
can be reached by applying the transformation $\hat{\Gamma}$ on a
$D4-D0$ system.

Under the aforementioned transformation, a $D4-D0$ system
transforms into $\hat{\Gamma}Q_{40}$
\begin{displaymath}
Q_{40}=\left(%
\begin{array}{c}
  0 \\
  p^1\\
  0 \\
  q_1\\
\end{array}%
\right)\rightarrow \hat{\Gamma}Q_{40}=\left(%
\begin{array}{c}
  -c(3d^2 p^1+c^2 q_0) \\
  d(2bc+ad)p^1+ac^2 q_0\\
  3(b(bc+2ad)p^1+a^2c q_0)\\
  a(3b^2p^1+a^2q_0)\\
\end{array}%
\right)
\end{displaymath}
The solution of the twistor $z$ with the new charges
$\hat{\Gamma}Q_{40}$ is
\begin{equation}
z=\frac{a\pm \sqrt{a^2+c^2}}{c}
\end{equation}
independent of the $D4-D0$ charges we started with. Now given an
arbitrary charge $Q$, there exists a transformation
$\hat{\Gamma}_{Q}$ such that $Q=\hat{\Gamma}_{Q}Q_{40}$ for some
$Q_{40}$. The twistor $z$ remains invariant under $Q\rightarrow
Q\tau +\hat{\Gamma}_{Q}h_{40}$ for arbitrary $h_{40}$. We conclude
that the non-BPS single-centered black holes that can be
constructed via the naive harmonic function procedure are only
those with $(Q,h)$ being the image of a single transformation
$\hat{\Gamma}$ on the $(Q_{40},h_{40})$ from a $D4-D0$ system:
\begin{equation}\label{QharmonincProcedureNB}
x_{NB}(\tau)=x^{*}_{NB}(\hat{\Gamma} Q_{40}\rightarrow
\hat{\Gamma}Q_{40}\tau+\hat{\Gamma}h_{40})\qquad
y_{NB}(\tau)=y^{*}_{NB}(\hat{\Gamma} Q_{40}\rightarrow
\hat{\Gamma}Q_{40}\tau+\hat{\Gamma}h_{40})
\end{equation}
Since we are considering arbitrary charge system, the constraint
is on the allowed values of $h$.

\section{Multi-centered Attractor Flows in $G_{2(2)}/(SL(2,\mathbb{R})\times SL(2,\mathbb{R}))$ model}

As proven in the pure gravity system, the multi-centered attractor
solutions are given by exponentiating the matrix harmonic function
$K(\vec{x})$:
\begin{equation}
 S(\vec{x})=e^{K(\vec{x})}S_0
\end{equation}
with $K(\vec{x})$ having the same properties as the generator $k$:
\begin{equation}
K^3(\vec{x})=0 \qquad \textrm{and} \qquad K^2(\vec{x})
\,\,\textrm{rank two}
\end{equation}
We now describe how to formulate $K(\vec{x})$ for multi-centered solutions in $G_{2(2)}$.

The $K(\vec{x})$ satisfying all the above constraints is
constructed as
\begin{equation}
K(\vec{x})=\sum_{a=1,2} [v_a W_a(\vec{x})^T+ W_a(\vec{x})
v_a^T]S_0
\end{equation}
with $v_a$ being the same two constant null vectors in $k$, and
the multi-centered harmonic function
\begin{equation}
W_a(\vec{x})=\sum_i\frac{(w_a)_i}{|\vec{x}-\vec{x}_i|}+m_a
\end{equation}
is everywhere orthogonal to $v_a$. The two 7-vectors
$(m_1,m_2)$ contain the information of asymptotic moduli and has
the same form as $(w_1,w_2)$. Write $K(\vec{x})$ as
$K(\vec{x})=\sum_i\frac{k_i}{|\vec{x}-\vec{x}_i|}+g$ where
\begin{equation}
k_i=\sum_{a=1,2} [v_a (w_a)_i^T+ (w_a)_i v_a^T]S_0  \qquad
\textrm{and} \qquad g=\sum_{a=1,2}[v_a m_a^T+ m_a v_a^T] S_0
\end{equation}
Since $v$ only depends on the twistor $(z,u)$, and $(w_1, w_2)$
are linear in $C^A$ or $C^{\alpha a}$, the above generating
procedure is equivalent to replace $C^{A}$ or $C^{\alpha a}$ by
the multi-centered harmonic functions while keeping the twistor
$(z,u)$ fixed.

Next we discuss the properties of the BPS multi-centered attractor
solution and non-BPS ones separately, since they are very different
in character.

\subsection{BPS Multi-centered Solutions}

In constrast with the multi-centered solutions in pure gravity, now
the second term of the current $J=\nabla K+ \frac{1}{2}[\nabla K,
K]$ does not vanish automatically since
\begin{equation}
[k^{BPS}_i,k^{BPS}_j] \neq 0 \qquad \textrm{and} \qquad
[k^{BPS}_i,g^{BPS}] \neq 0
\end{equation}
Therefore, the centers are no longer free, and we cannot simply
read off the charges from $J$. Instead, we need to solve for
$C^A_i$ and $G^A$ in a set of $5N$ coupled equations. The
divergence of the current is
\begin{equation}
\nabla \cdot J=4 \pi
\sum_i\delta(\vec{x}-\vec{x}_i)S_0(k_i+\frac{1}{2}[k_i,
g]+\frac{1}{2}\sum_j\frac{[k_i,k_j]}{|\vec{x}_i-\vec{x}_j|})S_0
\end{equation}
Using $\mathbf{Q}_i$ to denote the charge matrix which relates to
the D-brane charge $\{p^0,p^1,q_1,q_0\}_i$ as in (\ref{QDbranes}),
and with $Q_{32}$ as the vanishing NUT charge, we have $5N$ coupled
equations from $\mathbf{Q}_i=\frac{1}{4\pi}\int_i \nabla \cdot J$:
\begin{equation}\label{kfromQmulti}
\mathbf{Q}_i=S_0 (k_i+\frac{1}{2}[k_i,
g]+\frac{1}{2}\sum_j\frac{[k_i,k_j]}{|\vec{x}_i-\vec{x}_j|})S_0
\end{equation}

The generators of the multi-centered BPS attractor solution
$\{k_i\}$ and $g$ have $4(N+1)+2$ parameters in total: the two
twistors $(z,u)$ and $\{C^{A}_i, G^A\}$. On the left hand side of
(\ref{kfromQmulti}), there are also $3N-3$ degrees of freedom from
the position of the centers $\vec{x}_i$. On the other hand, a
generic $N$-centered attractor solution has $4N$ D-brane charges
$(p^I, q_I)$, and three additional constraints from the asymptotic
moduli $(x_0, y_0)$ and $u_0=1$. As we will show, like the
single-centered BPS solution, the three asymptotic moduli,
together with the vanishing of the total Taub-NUT charge,
determine the $4$ $G^A$ inside $g$. Moreover, as in the
single-centered case, we can solve $C^A_i$ in terms of the $4D$
D-brane charges $Q_i$ while leaving $(z,u)$ unfixed. The remaining
$N-1$ zero Taub-NUT charge conditions at each center will impose
$N-1$ constraints on the distances between the $N$-centers
$\vec{x}_i$.

First, integrating over the circle at the infinity, $\sum_i
\mathbf{Q}_i=\frac{1}{4\pi}\int\nabla \cdot J$ gives the sum of
the above $N$ matrix equations:
\begin{equation}\label{totalNUTeq}
\mathbf{Q}_{tot}=\sum_i \mathbf{Q}_i=S_0(\sum_i
k_i+\frac{1}{2}[\sum_i k_i, g])S_0
\end{equation}
which is the same as the one for the single-center attractor with
charge $Q_{tot}$. This determines $g$ to be $g=g_h+\Lambda$, same
as the one for single-centered attractor in (\ref{gghlambda}). As
in the single centered case, $h$ is fixed by the asymptotic moduli
$(x_0, y_0)$ by
\begin{equation}
x_0= x_{BPS}^{*}(Q\rightarrow h) \qquad
y_0=y_{BPS}^{*}(Q\rightarrow h)
\end{equation}
and the two constraints:
\begin{equation}
<Q_{tot},h>=0 \qquad J_4(h^0,h^1,\frac{h_1}{3},h_0)=1
\end{equation}
We have used the vanishing of the total Taub-NUT charge to
determine $\Lambda$. Next, we will use the remaining coupled
$5N-1$ equations to solve for the $4N$ $\{C^A_i\}$ and impose $N-1$
constraints on the relative positions between the $N$ centers
$\{\vec{x}_i\}$ where  $i=1,\cdots, N$.

The tentative solutions of $C^A_i$ are given by
(\ref{CfromQzBgenericz}) with $(p^I,q_I)$ replaced by
$(p^I_i,q_{I,i})$. The flow generator of each center $k_i$ is then
$k_i=a_{\alpha A}z^{\alpha}C^{A}_i$. Substituting the solution of $k_i$
and $g=g_h+\Lambda$ into (\ref{kfromQmulti}), and using
\begin{equation}
[k_i, g_h]=2<Q_i, h> \Theta \qquad [k_i, k_j]=2<Q_i, Q_j> \Theta
\end{equation}
where all the $k_i$'s and $g_h$ have the same value for the twistor
$z$, we get
\begin{equation}
\mathbf{Q}_i=S_0 (k_i+<Q_i, h>\Theta+\frac{1}{2}[k_i,
\Lambda]+\sum_j\frac{<Q_i, Q_j>}{|\vec{x}_i-\vec{x}_j|}\Theta)S_0
\end{equation}
Just as in the single-centered case, the solution of $k_i$ and the
form of $\Lambda$ guarantee that
\begin{equation}
\mathbf{Q}_i=S_0 (k_i+\frac{1}{2}[k_i, \Lambda])S_0
\end{equation}
We see that as long as the following integrability condition is
satisfied:
\begin{equation}\label{integrabilitymulti}
<Q_i,h>+\sum_j\frac{<Q_i,Q_j>}{|\vec{x}_i-\vec{x}_j|}=0
\end{equation}
the $k_i$ and $g$ given above indeed produce the correct
multi-centered attractor solution. Just like in the single-centered
case, the multi-centered solution flows to the correct attractor
moduli $(x_i^{*},y^{*}_i)$ near each center, independent of the value
of $z$. It also follows that the multi-centered
solution can be generated by replacing the charges inside the
attractor value by the multi-centered harmonic function:
\begin{equation}
x_{BPS}(\vec{x})=x^{*}_{BPS}(Q\rightarrow
\sum_i\frac{Q_i}{|\vec{x}-\vec{x}_i|}+h) \qquad
y_{BPS}(\vec{x})=y^{*}_{BPS}(Q\rightarrow
\sum_i\frac{Q_i}{|\vec{x}-\vec{x}_i|}+h)
\end{equation}

The sum of the $N$ equations in the integrability condition
(\ref{integrabilitymulti}) reproduces the constraint on $h$:
$<Q_{tot},h>=0$. Thus the remaining $N-1$ equations impose $N-1$
constraints on the relative positions between the $N$ centers
$\{\vec{x}_i\}$ with $i=1,\cdots, N$.  From (\ref{DefinitionDualScalarOmega}) and (\ref{SigmaConservedCurrent}),
we see that
$\ast \hbox{d} \omega$ is given by $J_{23}$. Defining the angular momentum $\vec{J}$ by
\begin{equation}
\label{DefinitionAngularMomentum} \omega_i = 2\epsilon_{ijk} J^j
\frac{x^k}{r^3}  \qquad  \hbox{as} \ r\rightarrow \infty
\end{equation}
we see that there exists a nonzero
angular momentum given by
\begin{equation}
\vec{J}=\frac{1}{2}\sum_{i<j}\frac{\vec{x}_i-\vec{x}_j}{|\vec{x}_i-\vec{x}_j|}\langle
Q_i,Q_j\rangle
\end{equation}
Thus we have shown that our multi-centered BPS attractor solution
reproduces the one found in \cite{Bates:2003vx}.

\subsection{non-BPS Multi-centered Solutions.}

For given $(z,u)$ and $\{C^{\alpha a}_i,G^{\alpha a}\}$, the
non-BPS multi-centered solution is the same as the single-centered
one as in (\ref{abcdefNB}) with $H^{\alpha a}(\tau)$ replaced by
the multi-centered harmonic function $H^{\alpha a}(\vec{x})=\sum_i
\frac{C^{\alpha a}_i}{|\vec{x}-\vec{x}_i|}+G^{\alpha a}$
satisfying the constraint
\begin{equation}\label{NBCGconstraintmulti}
u=\frac{H^{22}_i(\vec{x})}{H^{12}_i(\vec{x})}=\frac{C^{22}_i}{C^{12}_i}=\frac{G^{22}}{G^{12}}
\end{equation}
Accordingly, the attractor values at each center is the same as
(\ref{moduli*CNBsingle}) with the corresponding $C^{\alpha a}_s$.
The asymptotic moduli are obtained by extraction from
$S=e^{g}S_0$.

The equation of motion for the non-BPS multi-centered solution
simplifies a great deal since
\begin{equation}
[\nabla K(\vec{x}),K(\vec{x})]=0
\end{equation}
automatically, following from the fact that for the non-BPS system:
\begin{equation}
(w_1)_i\cdot m_2=(w_2)_i\cdot m_1=0
\end{equation}
which are guaranteed by the forms of NonBPS $(w_1,w_2)_i$ and
$(m_1,m_2)$. Therefore, the $5N$ equations (\ref{kfromQmulti})
decouple into $N$ sets of $5$ coupled equations:
\begin{equation}\label{kfromQmultiNB}
\mathbf{Q}_i=S_0 (k_i)S_0
\end{equation}

Equation (\ref{kfromQmultiNB}) differs greatly from the BPS counterpart
(\ref{kfromQmulti}). Firstly, the generators of the multi-centered
non-BPS attractor solution $\{k_i\}$ and $g$ have $3(N+1)+2$
parameters: the two twistors $(z,u)$ and $\{C^{\alpha a}_i,
G^{\alpha a}\}$ with the constraint (\ref{NBCGconstraintmulti}).
In constrast to the BPS case, $g$ does not enter the equation. Thus
we can simply use the three asymptotic moduli, without invoking
the zero Taub-NUT condition, to determine the $3$ $G^{\alpha a}$
inside $g$. Secondly, unlike the BPS multi-centered solution, the
position of the centers $\vec{x}_i$ do not appear in the equation,
therefore there will be no constraint imposed on them: the centers
are free. Last but not least, the remaining $3N+2$ parameters in
$(z,u)$ and $C^{\alpha a}$ are not enough to parameterize a
generic $N$-centered attractor solution, which has $4N$ D-brane
charges $(p^I_i, q_{I,i})$. Accordingly, the multi-centered
non-BPS attractor generated by this ansatz will not have arbitrary
charges. Combining with the fact that $\vec{x}_i$ do not appear in
the R.H.S of the equation, we find that all the $N$ vanishing Taub-NUT
charge conditions can only act on the charges on the L.H.S. We
conclude that, in total, there will be $2N-2$ constraints on the
allowed charges.

Now we will show in detail the derivation of the constraints.
First, like in the single-centered NonBPS solution, the absence of
the Taub-NUT charge at infinity fixes $z$ via
\begin{equation}
\sum_i \mathbf{Q}_i=S_0(\sum_i k_i)S_0
\end{equation}
The solution is the same as the solution to (\ref{zfromQNB}) with
the charges replaced by the total charges of $N$ centers:
$z=z(Q\rightarrow \sum_i Q_i)$. Since all the $N$ centers share
the same twistor $z$, the absence of the NUT charge at each center
imposes $N-1$ constraints on the allowed charges $Q_i$: all
$z(Q_i)$ have to be equal.

The remaining $4N$ equations in (\ref{kfromQmultiNB}) determine
$C^{\alpha a}$ in terms of $z$ and $Q_i$. Since the N-centers
decouple, (\ref{kfromQmultiNB}) for each center is the same as the
single-center one (\ref{kfromQsingleNB}). Thus the solution of
$C^{\alpha a}_i$ is given by (\ref{CfromQzNB}) with $(p^I, q_I)$
replaced by $(p^I_i, q_{I,i})$. Again, since all the centers share
the same twistor $u$, the condition (\ref{NBCGconstraintmulti})
imposes another $N-1$ constraints on the allowed charges. Solving
these $2N-2$ constraints, we see all the charges $\{Q_i\}$ are the
image of a single transformation $\hat{\Gamma}$ on a
multi-centered $D4-D0$ system $Q_{40,i}$:
\begin{equation}
Q_{i}=\hat{\Gamma}Q_{40, i}
\end{equation}
The charges at different centers are all mutually local
\begin{equation}
\langle Q_i, Q_j \rangle =0
\end{equation}
Except for the constraint on the charges, the $N$ centers are
independent, and there is no constraint on the position of the
centers. A related fact is that the the angular momentum is zero.

Like the non-BPS single-centered case, though the multi-centered
solution can be generated from the attractor value by replacing
$C^{\alpha a}$ with the multi-centered harmonic function
$H^{\alpha a}(\vec{x})=\sum_i \frac{C^{\alpha
a}_i}{|\vec{x}-\vec{x}_i|}+G^{\alpha a}$ under the constraint
(\ref{NBCGconstraintmulti}), while leaving $z$ unchanged as in
(\ref{Charmonic}), the generic solution cannot be generated via
the harmonic function procedure used in the BPS case, namely, by
replacing the charges inside the attractor value by the
corresponding multi-centered harmonic functions. The reason is
again due to the fact that the twistor $z$, being a function of
charges, does not remain invariant under this substitution of
charges by harmonic functions. The multi-centered non-BPS
solutions that can be generated by the harmonic function procedure
are those with $\{Q_i, h\}$ being the image of a single
$\hat{\Gamma}$ on the $\{Q_{40,i},h_{40}\}$ of a pure $D4-D0$
system:
\begin{equation}\label{QharmonincProcedureNBmulti}
x_{NB}(\vec{x})=x^{*}_{NB}(\hat{\Gamma} Q_{40}\rightarrow
\sum_i\frac{\hat{\Gamma}Q_{40,i}}{|\vec{x}-\vec{x}_i|}+\hat{\Gamma}h_{40})\qquad
y_{NB}(\vec{x})=y^{*}_{NB}(\hat{\Gamma} Q_{40}\rightarrow
\sum_i\frac{\hat{\Gamma}Q_{40,i}}{|\vec{x}-\vec{x}_i|}+\hat{\Gamma}h_{40})
\end{equation}
It appears that the existence of a simple linear ansatz for
``superimposing" single center solutions exists in general only
for mutually local extremal black holes, and only in the
supersymmetric case does it extend to mutually non-local centers.

To summarize, the non-BPS multi-centered solution is different
from the BPS case because it imposes no constraints on the
position of the centers, but instead on the allowed charges $Q_i$:
the choice of charges at each center are restricted to a
three-dimensional subspace, and they are mutually local. The
result is that the centers can move freely, and there is no
angular momentum in the system. It does not have interesting
moduli spaces of centers with mutually non-local charges, so it is
as ``boring" as the pure gravity case.

\section{Conclusion and Discussion}
In this paper, we find exact single-centered and
multi-centered black hole solutions in theories of gravity which have a
symmetric 3D moduli space. The BPS and extremal non-BPS single-centered solutions correspond to certain geodesics in the moduli space. We construct these geodesics by exponentiating different types of nilpotent elements in the coset algebra. Using the Jordan form of these nilpotent elements, we are able to write them down in closed explicit form. Furthermore, we can use a symmetric matrix parametrization to recover the metric and full flow of the scalars in four dimensions.

We have also generalized the geodesics to find solutions for
non-BPS and BPS multi-centered black holes.  The BPS
multi-centered solution reproduces the known solution of Bates and
Denef. Given our assumption that the 3D spatial slice is flat, we
find that a non-BPS multi-centered black hole is very different
from its BPS counterpart. It is constrained to have mutually local
charges at all of its centers and therefore carries no intrinsic
angular momentum. It is possible that if we dropped this
assumption, we could find more general non-BPS multi-centered
solutions. Such configurations would probably be amenable to exact
analysis only in the axially symmetric case, using inverse
scattering methods.

There are many other avenues for future work. One could explore
nilpotent elements in other symmetric spaces, and see whether
non-BPS bound states with nonlocal charges exist. In particular,
it would be interesting to study $E_{8(8)}/SO^{*}(16)$, which is
the 3d moduli space for $d=4, {\cal N}=8$ supergravity. We would
also like to find a way to modify our method so that we can apply
it to non-symmetric homogeneous spaces, and eventually to generic
moduli spaces. We could then study the much larger class of
non-BPS extremal black holes in generic ${\cal N}=2$
supergravities.

\section*{Acknowledgements}

We thank A. Neitzke, J. Lapan and A. Strominger for helpful discussions. The
work is supported by DOE grant DE-FG02-91ER40654. M.P. is also
supported by an NSF Graduate Fellowship.

\appendix

\section{Derivation of the moduli space $\mathcal{M}_{3D}$}

Here we briefly review the derivation of the $3d$ moduli space
$\mathcal{M}_{3D}$ from the $c^{*}$-map of the $4D$ supergravity
coupled to $n_V$ vector multiplets \cite{de Wit:1992wf}
\cite{Ceresole:1995ca} \cite{Pioline:2006ni}.

The bosonic part of the action for the $\mathcal{N}=2$
supergravity coupled to $n_V$ vector-multiplets is:
\begin{equation}
S=-\frac{1}{16\pi}\int d^4x
\sqrt{g^{(4)}}\left[R-2g_{i\bar{j}}dz^i\wedge \ast_4
d\bar{z}^{\bar{j}}-F^I\wedge G_I\right]
\end{equation}
where the ranges of the indices are $i,j=1,\dots,n_V$ and
$I=0,1,\dots, n_V$, and $G_I =  (Re{\cal N})_{IJ} F^J+(Im{\cal
N})_{IJ} \ast F^J$. The complex symmetric matrix
$\mathcal{N}_{IJ}$ is defined by
\begin{equation}
F_I=\mathcal{N}_{IJ}X^J \qquad \qquad
D_iF_I=\overline{\mathcal{N}}_{IJ}D_iX^J
\end{equation}
For model endowed with a prepotential $F(X)$,
\begin{equation}
\mathcal{N}_{IJ}=F_{IJ}+2i\frac{(\textrm{Im} F\cdot
X)_I(\textrm{Im} F \cdot X)_J}{X \cdot \textrm{Im} F \cdot X}
\end{equation}
where $F_{IJ}=\partial_I \partial_J F(X)$.

After reduction on the time-like isometry, the action is
$S=-\frac{1}{8\pi}\int dt\int d^3\textbf{x} \
\boldsymbol{\mathcal{L}}$. The 3D lagrangian
$\boldsymbol{\mathcal{L}}$ has three parts:
$\boldsymbol{\mathcal{L}}=\boldsymbol{\mathcal{L}}_{gravity}+\boldsymbol{\mathcal{L}}_{moduli}+\boldsymbol{\mathcal{L}}_{e.m}$
where
\begin{eqnarray}
\boldsymbol{\mathcal{L}}_{gravity} &=&  -
\frac{1}{2}\sqrt{\textbf{g}} \ \textbf{R}+\textbf{d}U \wedge
\boldsymbol{\ast} \textbf{d}U - \frac{1}{4} e^{4U}
\textbf{d}\omega \wedge \boldsymbol{\ast}
\textbf{d}\omega\non\\
\boldsymbol{\mathcal{L}}_{moduli}&=&g_{i\bar{j}}\textbf{d}z^i\wedge
\boldsymbol{\ast} \textbf{d}\bar{z}^{\bar{j}}\\
\boldsymbol{\mathcal{L}}_{e.m.}&=&\frac{1}{2}e^{-2U}(Im
\mathcal{N})_{IJ}\textbf{d}A^I_{0}\wedge
\boldsymbol{\ast}\textbf{d}A^J_{0}+\frac{1}{2}e^{2U}(Im
\mathcal{N})_{IJ}(\textbf{d}\textbf{A}^I+A^I_{0}\textbf{d}\omega)\wedge
\boldsymbol{\ast}(\textbf{d}\textbf{A}^J+A^J_{0}\textbf{d}\omega)\non\\
&&+(Re \mathcal{N})_{IJ}\textbf{d}A^I_0\wedge
(\textbf{d}\textbf{A}^J+A^J_{0}\textbf{d}\omega)\non
\end{eqnarray}
The dual scalars for $\omega$ and $\textbf{A}^I$ are defined by:
\begin{eqnarray}
e^{2U}(Im\mathcal{N})_{IJ}\boldsymbol{\ast}(\textbf{d}\textbf{A}^J+A^J_{0}\textbf{d}\omega)+(Re\mathcal{N})_{IJ}\textbf{d}A^J_0&=&-\textbf{d}\phi_{\textbf{A}^I} \non\\
\label{DefinitionDualScalarOmega}
e^{4U}\boldsymbol{\ast}\textbf{d}\omega+(A^I_{0}\textbf{d}\phi_{\textbf{A}^I}-\phi_{\textbf{A}^I}\textbf{d}A^I_{0})&=&-\textbf{d}\phi_{\omega}
\end{eqnarray}
After renaming the variables  $\phi_{\omega} \rightarrow \sigma,
A^I_0 \rightarrow A^I, \phi_{\textbf{A}^I} \rightarrow B_I$, we
obtain the 3d lagrangian in terms of scalars only:
\begin{eqnarray}
\mathcal{L}&=&-\frac{1}{2}\sqrt{\textbf{g}}\
\textbf{R}+\textbf{d}U \wedge \boldsymbol{\ast} \textbf{d}U
+\frac{1}{4}e^{-4U}(\textbf{d}\sigma+A^I \textbf{d}B_I-B_I
\textbf{d}A^I)\wedge \boldsymbol{\ast}(\textbf{d}\sigma+A^I
\textbf{d}B_I-B_I \textbf{d}A^I)\non\\
&&+ g_{i\bar{j}}(z,\bar{z})\textbf{d}z^i \wedge \boldsymbol{\ast}
\textbf{d}\bar{z}^{\bar{j}}+
\frac{1}{2}e^{-2U}(Im\mathcal{N})_{IJ}\textbf{d}A^I \wedge
\boldsymbol{\ast}
\textbf{d}A^J\non\\
&&+\frac{1}{2}e^{-2U}(Im\mathcal{N}^{-1})^{IJ}(\textbf{d}B_I+(Re\mathcal{N})_{IK}\textbf{d}A^K)\wedge
\boldsymbol{\ast}(\textbf{d}B_J+(Re\mathcal{N})_{JL}\textbf{d}A^L) \non\\
&=&-\frac{1}{2}\sqrt{\textbf{g}} \ \textbf{R}+g_{mn}\partial_a
\phi^m
\partial^a \phi^n
\end{eqnarray}
where, as before,  $\phi^n$ are the $4(n_V+1)$ moduli fields:
$\phi^n=\{U,z^i,\bar{z}^{\bar{i}},\sigma,A^I,B_I\}$, and
$\textbf{g}_{ab}$ is the space time metric, $g_{mn}$ is the moduli
space metric. Therefore, the moduli space $\mathcal{M}_{3D}$ has
metric:
\begin{eqnarray}
ds^2&=&\textbf{d}U \cdot \textbf{d}U
+\frac{1}{4}e^{-4U}(\textbf{d}\sigma+A^I \textbf{d}B_I-B_I
\textbf{d}A^I)\cdot (\textbf{d}\sigma+A^I \textbf{d}B_I-B_I
\textbf{d}A^I)+g_{i\bar{j}}(z,\bar{z})\textbf{d}z^i \cdot
\textbf{d}\bar{z}^{\bar{j}}\non\\
&&+\frac{1}{2}e^{-2U}[(Im\mathcal{N}^{-1})^{IJ}(\textbf{d}B_I+\mathcal{N}_{IK}\textbf{d}A^K)\cdot
(\textbf{d}B_J+\overline{\mathcal{N}}_{JL}\textbf{d}A^L) ]
\end{eqnarray}
It is a para-quaternionic-K\"ahler manifold. Since the holonomy is
reduced from $SO(4n_V+4))$ to $Sp(2,\mathbb{R})\times
Sp(2n_V+2,\mathbb{R})$, the vielbein has two indices $(\alpha,A)$
transforming under $Sp(2,\mathbb{R})$ and $Sp(2n_V+2,\mathbb{R})$,
respectively. The para-quaternionic vielbein is the analytical
continuation of the quaternionic vielbein computed in
\cite{Ferrara:1989ik}:
\begin{displaymath}
V^{\alpha A}=\left(%
\begin{array}{cc}
  iu & v \non\\
  e^a & iE^a \non\\
  -i\bar{E}^{\bar{a}} & \bar{e}^{\bar{a}} \non\\
  -\bar{v} & i\bar{u} \non\\
\end{array}%
\right)
\end{displaymath}
The 1-forms are defined as\begin{eqnarray}
u &\equiv & e^{\mathcal{K}/2-U}X^I(\textbf{d}B_I+\mathcal{N}_{IJ}dA^J)\non\\
e^a &\equiv & e^a_i \textbf{d}z^i\non\\
E^a &\equiv & e^{-U}e^a_i g^{i\bar{j}}e^{\mathcal{K}/2}\bar{D}_{\bar{j}}X^I(\textbf{d}B_I+\mathcal{N}_{IJ}\textbf{d}A^J)\non\\
v &\equiv & -\textbf{d}U+\frac{i}{2}e^{-2U}(\textbf{d}a+A^I
\textbf{d}B_I-B_I \textbf{d}A^I)
\end{eqnarray}
where $e^a_i$ is the veilbein of the 4D moduli space, and the bar
denotes complex conjugate. The line element is related to the
vielbein by
\begin{equation}
ds^2=-u \cdot \bar{u}+g_{a\bar{b}}e^a \cdot
\bar{e}^{\bar{b}}-g_{a\bar{b}}E^a \cdot \bar{E}^{\bar{b}}+v \cdot
\bar{v}=\epsilon_{\alpha\beta}\epsilon_{AB}V^{\alpha A}\otimes
V^{\beta B}
\end{equation}
where $\epsilon_{\alpha\beta}$ and $\epsilon_{AB}$ are the
anti-symmetric tensors invariant under $Sp(2,\mathbb{R})\cong
SL(2,\mathbb{R})$ and $Sp(2n_v+2,\mathbb{R})$.

The isometries of the $\mathcal{M}^{*}_{3D}$ descends from the
symmetry of the 4D system. The gauge symmetries in 4D gives the
shifting isometries of $\mathcal{M}^{*}_{3D}$:
\begin{eqnarray}
A^I &\longrightarrow& A^I +\Delta A^I \non\\
B_I &\longrightarrow& B_I +\Delta B_I \\
\sigma &\longrightarrow& \sigma+\Delta \sigma+\Delta B_I
A^I-\Delta A^I B_I\non
\end{eqnarray}
The conserved currents and charges are given by (\ref{ConservedChargesCurrents}) and the discussion thereafter.

\end{document}